\documentclass[onecolumn,draftcls]{IEEEtran}
\usepackage[T1]{fontenc}
\usepackage[latin9]{inputenc}
\usepackage{color}
\usepackage{array}
\usepackage{float}
\usepackage{amsmath}
\usepackage{amsthm}
\usepackage{amssymb}
\usepackage{graphicx}
\usepackage{setspace}
\usepackage[unicode=true,
 bookmarks=true,bookmarksnumbered=true,bookmarksopen=true,bookmarksopenlevel=1,
 breaklinks=false,pdfborder={0 0 0},pdfborderstyle={},backref=false,colorlinks=false]
 {hyperref}
\hypersetup{pdftitle={Your Title},
 pdfauthor={Your Name},
 pdfpagelayout=OneColumn, pdfnewwindow=true, pdfstartview=XYZ, plainpages=false}

\makeatletter

\providecommand{\tabularnewline}{\\}
\newcommand{\lyxdot}{.}

\floatstyle{ruled}
\newfloat{algorithm}{tbp}{loa}
\providecommand{\algorithmname}{Algorithm}
\floatname{algorithm}{\protect\algorithmname}

\theoremstyle{plain}
\newtheorem{lem}{\protect\lemmaname}
\theoremstyle{plain}
\newtheorem{thm}{\protect\theoremname}

\usepackage[caption=false,font=footnotesize]{subfig}
\usepackage{algorithmic}

\@ifundefined{showcaptionsetup}{}{%
 \PassOptionsToPackage{caption=false}{subfig}}
\usepackage{subfig}
\makeatother

\providecommand{\lemmaname}{Lemma}
\providecommand{\theoremname}{Theorem}

\begin{document}
\title{\singlespacing{}Fundamental Limits and Optimization of Multiband Sensing}
\author{\singlespacing{}{\normalsize{}Yubo Wan, }\textit{\normalsize{}Graduate Student Member,
IEEE}{\normalsize{}, An Liu, }\textit{\normalsize{}Senior Member,
IEEE}{\normalsize{}, Rui Du, and Tony Xiao Han,}\textit{\normalsize{}
Senior Member, IEEE}{\normalsize{}}\thanks{This work was supported in part by National Science Foundation of
China (No.62071416), and in part by Huawei Technologies Co., Ltd.
(Corresponding authors: An Liu.)

Yubo Wan and An Liu are with the College of Information Science and
Electronic Engineering, Zhejiang University, Hangzhou 310027, China
(email: wanyb@zju.edu.cn; anliu@zju.edu.cn).

Tony Xiao Han and Rui Du are with Huawei Technologies Co., Ltd. (email:
tony.hanxiao@huawei.com).}}

\maketitle
\vspace{-0.5in}

\begin{abstract}
Multiband sensing is a promising technology that utilizes multiple
non-contiguous frequency bands to achieve high-resolution target sensing.
\textcolor{blue}{However, rare studies have investigated the fundamental
limits and optimization of multiband sensing systems.} In this paper,
we investigate the fundamental limits and optimization of multiband
sensing, focusing on the fundamental limits associated with time delay.
We first derive a closed-form expression of the Cramér-Rao bound (CRB)
for the delay separation to reveal useful insights. Then, a metric
called the statistical resolution limit (SRL) that provides a resolution
limit is employed to investigate the fundamental limits of delay resolution.
The fundamental limits of delay estimation are also investigated based
on the CRB and Ziv-Zakai bound (ZZB). Based on the above derived fundamental
limits, numerical results are presented to analyze the effect of frequency
band apertures and phase distortions on the performance limits of
the multiband sensing systems. Inspired by the fundamental limits
analysis, we formulate an optimization problem to find the optimal
system configuration in multiband sensing systems with the objective
of minimizing the delay SRL. To solve this non-convex constrained
problem, we propose an efficient alternating optimization (AO) based
algorithm which iteratively optimizes the variables using successive
convex approximation (SCA) and one-dimensional search. Simulation
results demonstrate the effectiveness of the proposed algorithm.
\end{abstract}


\section{Introduction}

Using wireless systems for target sensing has sparked considerable
interests in the recent years, and it fosters a wide range of emerging
applications such as indoor localization \cite{DFPLocalization,DFPLocaTOA},
activity recognition \cite{WiFisensing1,WiFisensing2,DFPHumanDet},
and integrated sensing and communication (ISAC) \cite{ISACsens1,ISACsens2,ISAC_An},
etc. To achieve high-accuracy target sensing, these applications need
to rely on the channel state information (CSI), which reveals important
information about the multipath propagation environment.

\begin{figure}[t]
\centering{}\includegraphics[width=8cm]{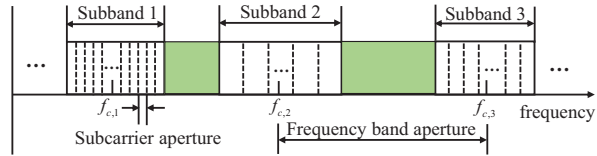}\caption{\label{fig:MBDistri}The frequency distribution of the multiband sensing
systems.}
\end{figure}

However, the target sensing performance is limited by the delay resolution,
which is inversely proportional to the bandwidth of the transmitted
signal. To address this issue, the multiband technology is proposed,
which provides the potential to achieve high-resolution target sensing
by making use of the CSI measurements across multiple non-contiguous
frequency bands. As shown in Fig. \ref{fig:MBDistri}, the spectrum
resource used for target sensing consists of a number of subbands
in the presence of frequency band apertures, where $f_{c,1}$, $f_{c,2}$,
and $f_{c,3}$ are carrier frequencies. \textcolor{blue}{The frequency
subbands allocated to other systems are illustrated by the green color,
which cannot be utilized for target sensing.} A few multiband based
sensing algorithms have been proposed recently, which achieve high-accuracy
multipath channel delay estimation for ranging and localization, and
illustrate that the improvement of estimation accuracy is brought
by the frequency band apertures \cite{nsdi,CS2019,CS2020,ESPRIT2,TSGE,LTESAGE}.

It is well known that fundamental limits not only serve as a performance
bound for practical multiband sensing technologies, but may also provide
useful guidance and insights for the design and analysis of multiband
sensing systems. However, the fundamental limits of a multiband based
sensing problem have not been fully investigated, especially under
the practical consideration of the phase distortions caused by hardware
imperfections \cite{nonidealfoctor3,nonidealfoctor1,nonidealfoctor2}.
Only a few studies have investigated the effect of frequency band
apertures on delay resolution and delay estimation accuracy based
on a fundamental limit analysis. In \cite{ESPRIT2,ConcateCRB1}, the
Cramér-Rao bound (CRB) is derived for the delay estimation error based
on a multiband signal model. However, there are several limitations:
(i) The authors only empirically showed that the CRB decreases with
the increase of frequency band apertures via numerical simulations
without a theoretical analysis; (ii) The effect of phase distortions
is not considered; (iii) The CRB is a local bound, which may not be
tight over a wide range of frequency band apertures. In \cite{SRL},
the authors derived the closed-form expression of the statistical
resolution limit (SRL) based on a simple pole model and showed the
effect of band apertures on the resolution. The SRL is defined as
the source separation that equals its own root squared CRB, which
provides a performance bound on the resolution of any practical method.
\textcolor{blue}{However, the derived results are restricted to a
simple pole model and have approximation errors}. Hence, in the aforementioned
studies, detailed effects of frequency band apertures and phase distortions
on the fundamental limits of delay estimation and delay resolution
remain underexplored.

Besides, determining how to improve the delay resolution limit by
designing the system parameters (e.g., the carrier frequency and the
number of subcarriers for each subband) in multiband sensing systems
is another challenging problem. To the best of our knowledge, there
is little literature refers to designing the multiband sensing system
parameters at the transmitter for the purpose of improving the sensing
performance limits. In \cite{SelectionCRB2}, the authors proposed
a sparse subbands selection methodology for ranging based on the CRB.
However, it only involves subbands selection with fixed system parameters.
In addition, the path gain variables are assumed to be real, which
is restrictive.\textcolor{blue}{{} In \cite{B_subarray}, the authors
investigated a problem of multiple subarrays geometry design, which
is similar to the optimization problem of multiband sensing. However,
they only considered a simplified optimization problem of minimizing
the CRB for a single source DOA estimation and did not investigate
the fundamental limits in terms of SRL or ZZB.}

In this paper, we study the effect of frequency band apertures and
phase distortions on the multiband target sensing systems based on
a comprehensive fundamental limit analysis, which is the key in understanding
the multiband sensing systems. The Fisher information matrix (FIM)
based on a practical multiband signal model is derived first, where
the signal model considers the phase distortions, e.g., receiver timing
offset and random phase offset. Then, we reformulate the FIM into
a compact form using the Dirichlet kernel \cite{DirichletKernel}
to derive a closed-form expression of the CRB for the delay separation
in a simplified case to reveal useful insights. Besides, numerical
analyses are presented employing the metric SRL, that provides a fundamental
limit of delay resolution in the presence of phase distortions. The
performance bounds CRB and Ziv-Zakai bound (ZZB) are derived to analyze
the fundamental limits of delay estimation, where the ZZB is a global
bound computed by transforming the estimation problem to a binary
hypothesis testing problem \cite{ZZB1,ZZB2,ZZB3}.\textcolor{blue}{{}
From the fundamental limits analysis, we have the following key insights:
(1) The CRB for delay separation decreases with the increase of frequency
band apertures in a square order; (2) Increasing the frequency band
apertures leads to an improvement of delay resolution limit and the
phase distortions have relatively slight interference on SRL; (3)
The phase distortions degrade the performance limits of delay estimation
and make it difficult to exploit the delay estimation performance
gain brought by the frequency band apertures. However, when the targets
are distinguishable with significantly different time delay and amplitudes,
the negative effect of phase distortions can be suppressd; (4) ZZB
provides a tighter bound than CRB over a wide range of frequency band
apertures. Furthermore, the ZZB predicts a threshold behavior of a
maximum a posteriori (MAP) estimator, i.e., the ZZB decreases first
and then increases with the increase of frequency band apertures,
which is consistent with the MAP estimation results.}

Inspired from the above analyses, an optimization problem is formulated
to find the optimal system configuration in multiband sensing systems,
where the carrier frequency and the subcarrier number of each subband
are jointly designed to minimize the delay SRL under the total bandwidth
constraint. Besides, the carrier frequencies of all subbands are respectively
constrained within a certain frequency interval to reflect the practical
constraints that only a few non-contiguous subbands are available
for target sensing. The formulated problem is difficult to solve since
it has a non-convex equality constraint associated to the definition
of SRL. Furthermore, the optimization variable, subcarrier number,
is integer with the form of summation terms in the CRB expression,
which is difficult to optimize. To overcome these challenges, we relax
the integer variable to a real variable and employ the Dirichlet kernel
to compactly reformulate the summation terms. Then, we adopt alternating
optimization (AO) algorithms to alternatively optimize the variables
of delay separation and system parameters. For given system parameters,
the optimal delay SRL can be found by one-dimensional search of the
delay separation. For given delay separation, the system parameters
are optimized to minimize the CRB of the delay separation. We adopt
the successive convex approximation (SCA) algorithm to solve this
non-convex subproblem. \textcolor{blue}{Finally, the convergence of
the overall algorithm has been proved.} Numerical results are provided
to validate the effectiveness of our proposed algorithms and present
useful guidance for the system design.

The rest of this paper is organized as follows. Section \ref{sec:System Model}
presents the system model and the derivations of the FIM. In Section
\ref{sec:Fundamental Limit of Delay Resolution}, we derive a closed-form
expression of the CRB for delay separation. Then we introduce the
SRL and analyze the fundamental limits of delay resolution. Section
\ref{sec:Fundamental Limit of Delay Estimation} studies the fundamental
limits of delay estimation based on the derivation of the CRB and
ZZB. Section \ref{sec:The Multiband Optimization Problem} formulates
a multiband sensing optimization problem and an efficient algorithm
is proposed based on AO method. Finally, Section \ref{sec:Conclusion}
concludes the paper.

\textit{Notations}: $\mathbf{I}$ denotes an identity matrix, $\propto$
denotes equality up to irrelevant constants, $\delta\left(\cdot\right)$
denotes the Dirac's delta function, $\mathrm{diag}\left(\cdot\right)$
constructs a diagonal matrix from its vector argument, and $\|\cdot\|$
denotes the Euclidean norm of a complex vector. For a matrix $A$,
$A^{T},A^{H},A^{-1}$ represent a transpose, complex conjugate transpose,
and inverse of a matrix, respectively. $\mathbb{E}_{\mathbf{z}}[\cdot]$
denotes the expectation operator with respect to the random vector
$\mathbf{z}$. The notations $\mathbb{R}^{\textrm{+}}$ represents
the strictly positive real number and $\mathcal{CN}(\mathrm{\boldsymbol{\mu}},\boldsymbol{\Sigma})$
denotes a complex Gaussian normal distribution with mean $\boldsymbol{\mu}$
and covariance matrix $\boldsymbol{\Sigma}$.

\section{System Model and FIM\label{sec:System Model}}

\subsection{System Model}

\begin{figure}[tbh]
\centering{}\includegraphics[width=8cm]{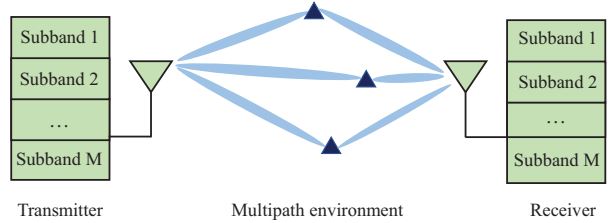}\caption{\label{fig:The-multiband-OFDM}An illustration of the multiband target
sensing system.}
\end{figure}

As shown in Fig. \ref{fig:The-multiband-OFDM}, we consider a single-input
single-output (SISO) multiband target sensing system which employs
orthogonal frequency division multiplexing (OFDM) training signals
over $M$ frequency subbands. Note that the \textcolor{blue}{multiband}
sensing is mainly used to improve the delay resolution and a SISO
system is sufficient to reveal the key insights on the fundamental
limits of \textcolor{blue}{multiband} sensing in terms of delay resolution/estimation.
Then, the continuous-time channel impulse response (CIR) $h\left(t\right)$
can be written as \cite{DFPHumanDet}
\begin{equation}
h(t)=\sum_{k=1}^{K}\alpha_{k}\delta\left(t-\tau_{k}\right),
\end{equation}
where $K$ denotes the number of multipath caused by the scatters
(targets) in the sensing environment between the transmitter and the
receiver, $\alpha_{k}\in\mathbb{C}$ denotes a complex scalar carrying
the amplitude and phase information of the $k$-th scatter, and $\tau_{k}\in\mathbb{R^{\textrm{+}}}$
denotes the time delay carrying the range information of the $k$-th
scatter. The delays are sorted in an increasing order, i.e., $\tau_{k-1}<\tau_{k}$,
$k=2,...,K$. As in \cite{CS2019,CS2020,ESPRIT2}, we assume that
$\alpha_{k},\forall k$ and $\tau_{k},\forall k$ are independent
of the frequency subbands, and denote that each subband has $N_{m}$
orthogonal subcarriers with subcarrier spacing $f_{s,m}$. The carrier
frequency of the $m$-th subband is denoted as $f_{c,m}$. Then, via
a Fourier transform of the CIR as in \cite{ESPRIT2}, the channel
frequency response (CFR) samples can be expressed as
\begin{equation}
\tilde{h}_{m,n}=\sum_{k=1}^{K}\alpha_{k}e^{-j2\pi f_{m,n}\tau_{k}},\label{eq:CFR}
\end{equation}
where $f_{m,n}=f_{c,m}+nf_{s,m}$, $m=1,...,M$, $n\in\mathcal{N_{\mathit{m}}}\triangleq\left\{ -\frac{N_{m}-1}{2},...,\frac{N_{m}-1}{2}\right\} $.
We assume that $N_{\mathit{m}},\forall m$ is an even number without
loss of generality, and denote $N=N_{1}+\ldots+N_{M}$ as the number
of CFR samples over all subbands. Then, during the period of a single
OFDM symbol, the discrete-time received signal model can be written
as \cite{CS2019,CS2020}
\begin{equation}
y_{m,n}\!=\!\sum_{k=1}^{K}\!\alpha_{k}e^{-j2\pi f_{m,n}\tau_{k}}e^{-j2\pi nf_{\!s,\!m}\delta_{m}}e^{j\varphi_{m}}s_{m,n}\!+\!w_{m,n},\label{eq:original_signal}
\end{equation}
where $w_{m,n}$ is the $n$-th element of the additive white Gaussian
noise (AWGN) vector $\boldsymbol{w}_{m}\in\mathbb{C}^{N_{m}\times1}$,
following the distribution $\mathcal{C}\mathcal{N}\left(0,\sigma_{ns}^{2}\mathbf{I}\right)$.
$s_{m,n}$ denotes a known training symbol over the $n$-th subcarrier
of the $m$-th subband with the power $\left|s_{m,n}\right|^{2}=1$.
The parameters $\varphi_{m}$ and $\delta_{m}$ represent the phase
distortions caused by random phase offset and receiver timing offset
\cite{nonidealfoctor3,nonidealfoctor1,nonidealfoctor2}, respectively.
In practice, the receiver timing offset $\delta_{m}$ is often within
a small range and thus we assume that $\delta_{m},\forall m$ follows
a prior distribution $p\left(\delta_{m}\right)\sim\mathcal{N}\left(0,\sigma_{p}^{2}\right)$,
where $\sigma_{p}$ is the timing synchronization error. \textcolor{blue}{Note
that the multiband system model has some similarities with the models
used in multiple subarrays signal processing, e.g., \cite{A_subarray}.
However, the signal model used in \cite{A_subarray} assumes that
each subarray is perfectly calibrated, while in our formulated multiband
signal model, each subband is not ``perfectly calibrated'' due to
the effect of $\delta_{m}$.}

The received signal model (\ref{eq:original_signal}) cannot be directly
used for the fundamental limit analysis due to the inherent ambiguity.
Specifically, for an arbitrary constant $c$, if we substitute two
sets of variables $(\left|\alpha_{k}\right|e^{j\angle\alpha_{k}},\varphi_{m})$
and $(\left|\alpha_{k}\right|e^{j(\angle\alpha_{k}+c)},\varphi_{m}-c)$
into (\ref{eq:original_signal}), the same observation result will
be obtained. It indicates that the parameters $(\alpha_{k},\varphi_{m})$
are ambiguous, which will result in a singular FIM \cite{FIM}. Therefore,
we equivalently transform the signal model (\ref{eq:original_signal})
by absorbing the phase $\varphi_{1}$ and center frequency $f_{c,1}$
of the first subband as \cite{TSGE}
\begin{equation}
y_{m,n}=\mu_{m,n}+w_{m,n},\label{eq:refined_signal}
\end{equation}
where
\[
\mu_{m,n}\!=\!\sum_{k=1}^{K}\!\alpha_{k}^{\prime}e^{-j2\pi f_{\!c,\!m}^{\prime}\!\tau_{k}}e^{-j2\pi nf_{\!s,\!m}\!\tau_{k}}e^{-j2\pi nf_{\!s,\!m}\delta_{m}}e^{j\varphi_{m}^{\prime}}s_{m,n},
\]
$f_{c,m}^{\prime}=f_{c,m}-f_{c,1},\alpha_{k}^{\prime}=\alpha_{k}e^{j\varphi_{1}}e^{-j2\pi f_{c,1}\tau_{k}}$,
and $\varphi_{m}^{\prime}=\varphi_{m}-\varphi_{1},\forall k,m$. The
new signal model (\ref{eq:refined_signal}) eliminates the inherent
ambiguity and reserves the structure of frequency band apertures,
i.e., $e^{-j2\pi f_{\!c,\!m}^{\prime}\!\tau_{k}}$. \textcolor{blue}{Note
that it is more convenient to define the frequency band aperture as
$f_{c,m}^{\prime}=f_{c,m}-f_{c,1}$ instead of the entire frequency
span because the fundamental limits such as CRB can be expressed as
a function of $f_{c,m}^{\prime}$'s. The proof of parameter identifiability
of signal model (5) can be found in Appendix \ref{subsec:parameter identifiability}.}

\subsection{FIM Derivation}

Let $\boldsymbol{\eta}=[\boldsymbol{\tau}^{T},\boldsymbol{\alpha}^{T},\boldsymbol{\varphi}^{T},\boldsymbol{\delta}^{T}]^{T}\in\mathbb{R}^{3K+2M-1}$
be the vector consisting of the unknown parameters, where
\begin{equation}
\begin{aligned} & \boldsymbol{\tau}=\left[\tau_{1},\tau_{2},\ldots,\tau_{K}\right]^{T},\\
 & \boldsymbol{\alpha}=\left[\boldsymbol{\alpha}_{R}^{T},\boldsymbol{\alpha}_{I}^{T}\right]^{T},\\
 & \boldsymbol{\alpha}_{R}=\left[\alpha_{R,1},\alpha_{R,2},\ldots,\alpha_{R,K}\right]^{T},\\
 & \boldsymbol{\alpha}_{I}=\left[\alpha_{I,1},\alpha_{I,2},\ldots,\alpha_{I,K}\right]^{T},\\
 & \boldsymbol{\varphi}=\left[\varphi_{2}^{\prime},\ldots,\varphi_{M}^{\prime}\right]^{T},\\
 & \boldsymbol{\delta}=\left[\delta_{1},\ldots,\delta_{M}\right]^{T},
\end{aligned}
\end{equation}
in which $\alpha_{R,k}$ and $\alpha_{I,k}$ denote the real and imaginary
parts of $\alpha_{k}^{\prime}$, respectively. Defining $\hat{\boldsymbol{\eta}}$
as the unbiased estimator of $\boldsymbol{\eta}$ based on the multiband
observations
\begin{equation}
\mathbf{y}=\left[y_{1,-\frac{N_{m}-1}{2}},\cdots,y_{M,\frac{N_{m}-1}{2}}\right]^{T}\in\mathbb{C}^{N\times1}.
\end{equation}
Then, the mean squared error (MSE) matrix of $\hat{\boldsymbol{\eta}}$
satisfies the information inequality \textcolor{blue}{\cite{LLR,BCRB}}
\begin{equation}
\mathbb{E}_{\boldsymbol{y},\boldsymbol{\delta}}\left[(\hat{\boldsymbol{\eta}}-\boldsymbol{\eta})(\hat{\boldsymbol{\eta}}-\boldsymbol{\eta})^{\mathrm{T}}\right]\succeq\mathbf{J}_{\boldsymbol{\eta}}^{-1},
\end{equation}
where $\mathbf{J}_{\boldsymbol{\eta}}$ denotes the $(3K+2M-1)\times(3K+2M-1)$
FIM with a priori knowledge of $\boldsymbol{\delta}$, defined as
\begin{equation}
\begin{aligned} & \mathbf{J}_{\boldsymbol{\eta}}=\mathbf{J}_{w}+\mathbf{J}_{p},\\
 & \mathbf{J}_{w}\triangleq\mathbb{E}_{\boldsymbol{y},\boldsymbol{\delta}}\left[-\frac{\partial^{2}\ln f(\boldsymbol{y}|\boldsymbol{\eta})}{\partial\boldsymbol{\eta}\partial\boldsymbol{\eta}^{T}}\right],\\
 & \mathbf{J}_{p}\triangleq\mathbb{E}_{\boldsymbol{\delta}}\left[-\frac{\partial^{2}\ln f(\boldsymbol{\delta})}{\partial\boldsymbol{\eta}\partial\boldsymbol{\eta}^{T}}\right],
\end{aligned}
\label{eq:FIM1}
\end{equation}
where $\mathbf{J}_{w}$ and $\mathbf{J}_{p}$ are the FIMs from the
observations and the a priori knowledge of $\boldsymbol{\delta}$,
respectively. $f(\mathbf{y}|\boldsymbol{\eta})\propto\exp\{-\frac{\|\mathbf{y}-\boldsymbol{\mu}\|_{2}^{2}}{\sigma_{ns}^{2}}\}$
is the likelihood function of the random vector $\mathbf{y}$ conditioned
on $\boldsymbol{\eta}$ and $f(\boldsymbol{\delta})\propto\exp\{-\frac{\|\boldsymbol{\delta}\|_{2}^{2}}{2\sigma_{p}^{2}}\}$
is the prior distribution of $\boldsymbol{\delta}$, where $\boldsymbol{\mu}=[\mu_{1,-(N_{m}-1)/2},\cdots,\mu_{M,(N_{m}-1)/2}]^{T}\in\mathbb{C}^{N\times1}$.
The FIM $\mathbf{J}_{\boldsymbol{\eta}}$ can be structured as
\begin{equation}
\mathbf{J}_{\boldsymbol{\eta}}=\left[\begin{array}{cccc}
\Psi(\boldsymbol{\tau},\boldsymbol{\tau}) & \Psi(\boldsymbol{\tau},\boldsymbol{\alpha}) & \Psi(\boldsymbol{\tau},\boldsymbol{\varphi}) & \Psi(\boldsymbol{\tau},\boldsymbol{\delta})\\
\Psi(\boldsymbol{\alpha},\boldsymbol{\tau}) & \Psi(\boldsymbol{\alpha},\boldsymbol{\alpha}) & \Psi(\boldsymbol{\alpha},\boldsymbol{\varphi}) & \Psi(\boldsymbol{\alpha},\boldsymbol{\delta})\\
\Psi(\boldsymbol{\varphi},\boldsymbol{\tau}) & \Psi(\boldsymbol{\varphi},\boldsymbol{\alpha}) & \Psi(\boldsymbol{\varphi},\boldsymbol{\varphi}) & \Psi(\boldsymbol{\varphi},\boldsymbol{\delta})\\
\Psi(\boldsymbol{\delta},\boldsymbol{\tau}) & \Psi(\boldsymbol{\delta},\boldsymbol{\alpha}) & \Psi(\boldsymbol{\delta},\boldsymbol{\varphi}) & \Psi(\boldsymbol{\delta},\boldsymbol{\delta})
\end{array}\right],\label{eq:FIM}
\end{equation}
where the entries of $\mathbf{J}_{\boldsymbol{\eta}}$ are derived
in Appendix \ref{subsec:FIM_ori}.

From (\ref{eq:FIM4}), we can observe that the FIM $\mathbf{J}_{\boldsymbol{\eta}}$
depends on the relative delay, e.g., $\tau_{2}-\tau_{1}$, rather
than on the absolute delay, which agrees with the results of \cite{SelectionCRB2},
but the phase distortions are not considered in their model. \textcolor{blue}{Besides,
though the FIM $\mathbf{J}_{\boldsymbol{\eta}}$ includes the entries
associated with phase distortion factors $\boldsymbol{\delta}$ and
$\boldsymbol{\varphi}$, it is independent of the value of $\boldsymbol{\delta}$
and $\boldsymbol{\varphi}$.}

\textcolor{blue}{For clarity, an overview of the main contents and
corresponding assumptions are summarized in Table \ref{tab:An-overview-of}.
The different assumptions adopted in different sections are just special
cases of the same general system model in (\ref{eq:refined_signal}).
The purpose of considering these special cases is to highlight the
impact of different system imperfections/parameters on the performance
and to reveal useful insight that cannot be easily obtain in the general
case.}

\textcolor{blue}{}
\begin{table*}[t]
\begin{centering}
\textcolor{blue}{\caption{\textcolor{blue}{\label{tab:An-overview-of}An overview of the main
contents and adopted basic assumptions in the paper.}}
}
\par\end{centering}
\centering{}\textcolor{blue}{}%
\begin{tabular}{|>{\centering}m{2cm}|>{\centering}m{2cm}|>{\centering}m{2cm}|>{\centering}m{2cm}|>{\centering}m{2cm}|>{\centering}m{2cm}|>{\centering}m{2cm}|}
\hline 
\textcolor{blue}{Section} & \multicolumn{1}{>{\centering}m{2cm}|}{\textcolor{blue}{\mbox{II}-B}} & \multicolumn{1}{>{\centering}m{2cm}|}{\textcolor{blue}{\mbox{III}-A}} & \textcolor{blue}{\mbox{III}-B} & \textcolor{blue}{\mbox{IV}-A} & \textcolor{blue}{\mbox{IV}-B} & \textcolor{blue}{\mbox{V}}\tabularnewline
\hline 
\textcolor{blue}{Contents} & \textcolor{blue}{FIM} & \textcolor{blue}{Closed-form CRB} & \textcolor{blue}{SRL} & \textcolor{blue}{DEB} & \textcolor{blue}{ZZB} & \textcolor{blue}{System optimization}\tabularnewline
\hline 
\textcolor{blue}{Basic}

\textcolor{blue}{Assumptions} & \textcolor{blue}{Exist phase distortions} & \textcolor{blue}{(i) Exist phase distortions}

\textcolor{blue}{(ii)$\alpha_{1}^{\prime}=\alpha_{2}^{\prime}=1$} & \textcolor{blue}{Exist phase distortions} & \textcolor{blue}{Exist phase distortions} & \textcolor{blue}{(i) Without phase distortions}

\textcolor{blue}{(ii) K=1} & \textcolor{blue}{(i) Exist phase distortions}

\textcolor{blue}{(ii) Coarse estimation of $\alpha_{k}$'s and $K$}\tabularnewline
\hline 
\end{tabular}
\end{table*}

\section{\label{sec:Fundamental Limit of Delay Resolution}Fundamental Limits
of Delay Resolution}

In this section, we consider a scenario with the number of subbands
$M=2$ and multipath number $K=2$, since the delay resolution reflects
the ability of resolving two paths with a small delay gap. Generally,
the derivation of CRB involves a complicated high-dimensional matrix
inverse operation. Hence, we derive a closed-form expression of the
CRB for the delay separation in a special case to reveal useful insight.
Finally, we investigate the fundamental limits of delay resolution
employing the metric SRL.

\subsection{\label{subsec:closed-form CRB}The Closed-form CRB for the Delay
Separation}

To derive the closed-form expression of CRB for the delay separation,
which is defined as $\Delta\tau=\left|\tau_{2}-\tau_{1}\right|$,
we need to reformulate the expression of FIM in (\ref{eq:FIM4}) into
a compact form by employing the Dirichlet kernel \cite{DirichletKernel}.
The details are elaborated in Appendix \ref{subsec:B.Compactly FIM}.

Subsequently, the CRB for the delay separation in the presence of
phase distortions is given by
\begin{eqnarray}
C_{\Delta\tau} & = & \frac{\partial\Delta\tau}{\partial\boldsymbol{\eta}}^{T}\mathbf{C}_{\boldsymbol{\eta}}\frac{\partial\Delta\tau}{\partial\boldsymbol{\eta}}\label{eq:C_delattau}\\
 & = & \mathbf{C}_{\boldsymbol{\eta}}(1,1)+\mathbf{C}_{\boldsymbol{\eta}}(2,2)-\mathbf{C}_{\boldsymbol{\eta}}(1,2)-\mathbf{C}_{\boldsymbol{\eta}}(2,1),\nonumber 
\end{eqnarray}
where $\mathbf{C}_{\boldsymbol{\eta}}=\mathbf{J}_{\boldsymbol{\eta}}^{-1}$
is the CRB associated with the vector $\boldsymbol{\eta}$. Particularly,
we assume that $\sigma_{ns}^{2}=2$, $f_{s,1}=f_{s,2}=f_{s}$, $N_{1}=N_{2}=\overline{N}$,
the amplitudes and phases of the two paths are set as $a_{1}=a_{2}=1$
and $\phi_{1}=\phi_{2}=0$ for simplicity. As can be seen, determining
$C_{\Delta\tau}$ requires inverting the high-dimensional FIM $\mathbf{J}_{\boldsymbol{\eta}}$,
while only a small submatrix $\left[\mathbf{J}_{\boldsymbol{\eta}}^{-1}\right]_{2\times2}$
is of interest. To circumvent high-dimensional matrix inversion, we
introduce the equivalent FIM (EFIM) \cite{wireless_locTOA4}. Given
parameters $\boldsymbol{\theta}_{1}=[\tau_{1},\tau_{2}]^{T}$ and
the FIM $\mathbf{J}_{\boldsymbol{\eta}}$ with the block matrix form
\begin{equation}
\mathbf{J}_{\boldsymbol{\eta}}=\left[\begin{array}{cc}
\mathbf{A} & \mathbf{B}\\
\mathbf{B}^{T} & \mathbf{C}
\end{array}\right],
\end{equation}
where $\mathbf{A}\in\mathbb{R}^{2\times2}$, $\mathbf{B}\in\mathbb{R}^{2\times7}$,
$\mathbf{C}\in\mathbb{R}^{7\times7}$, the EFIM can be written as
\begin{equation}
\mathbf{J}_{\mathrm{e}}\left(\boldsymbol{\theta}_{1}\right)\triangleq\mathbf{A}-\mathbf{B}\mathbf{C}^{-1}\mathbf{B}^{T}.
\end{equation}
Since $\left[\mathbf{J}_{\boldsymbol{\eta}}^{-1}\right]_{2\times2}=\mathbf{J}_{\mathrm{e}}^{-1}\left(\boldsymbol{\theta}_{1}\right)$,
(\ref{eq:C_delattau}) can be rewritten as
\begin{equation}
C_{\Delta\tau}=\mathbf{J}_{\mathrm{e}}^{-1}\left(1,1\right)+\mathbf{J}_{\mathrm{e}}^{-1}\left(2,2\right)-\mathbf{J}_{\mathrm{e}}^{-1}\left(1,2\right)-\mathbf{J}_{\mathrm{e}}^{-1}\left(2,1\right).
\end{equation}
Finally, we can obtain a closed-form expression of $C_{\Delta\tau}$
using symbolic algebra packages. However, it is difficult to directly
obtain insights from the closed-form expression due to its complicated
structure. Hence, we turn to obtain the tight lower and upper bound
of $C_{\Delta\tau}$, which are given by
\begin{equation}
\begin{aligned}\textrm{CRB}_{\textrm{up}} & =\frac{3\overline{N}+3\gamma}{(\overline{N}+\gamma)(3\overline{N}-3\gamma)\pi^{2}\mathit{\Delta f_{c}}^{2}+c},\\
\textrm{CRB}_{\textrm{low}} & =\frac{3\overline{N}}{\pi^{2}((3\overline{N}^{2}-3\gamma^{2})\mathit{\Delta f_{c}}^{2}+\overline{N}^{4}\mathit{fs}^{2}-\overline{N}^{2}\mathit{fs}^{2})},
\end{aligned}
\label{eq:CRBup_low}
\end{equation}
where $\Delta f_{c}=f_{c,2}-f_{c,1}$ denotes the frequency band apertures,
$c=(N+\gamma)(\overline{N}^{3}\mathit{fs}^{2}-\overline{N}\mathit{fs}^{2})\pi^{2}+3\overline{N}\gamma^{\prime\prime}+3\gamma\gamma^{\prime\prime}-3(\gamma^{\prime})^{2}$
is a constant coefficient independent of $\Delta f_{c}$, and the
definition of $\gamma$, $\gamma^{\prime}$, and $\gamma^{\prime\prime}$
can be seen in Appendix \ref{subsec:B.Compactly FIM}. \textcolor{blue}{The
derivation for $\textrm{CRB}_{\textrm{up}}$ and $\textrm{CRB}_{\textrm{low}}$
are presented in Appendix \ref{subsec:CRB_Closed_APP}}. Fig. \ref{fig:CRB_Closedform}
illustrates the square root of $C_{\Delta\tau}$, $\textrm{CRB}_{\textrm{up}}$,
and $\textrm{CRB}_{\textrm{low}}$ versus the delay separation $\Delta\tau$.
As can be seen, $\textrm{CRB}_{\textrm{up}}$ and $\textrm{CRB}_{\textrm{low}}$
are tight enough to force $C_{\Delta\tau}$, which indicates that
$C_{\Delta\tau}$ has similar properties with $\textrm{CRB}_{\textrm{up}}$
and $\textrm{CRB}_{\textrm{low}}$. From (\ref{eq:CRBup_low}), we
conclude that $\textrm{CRB}_{\textrm{up}}$ ($\textrm{CRB}_{\textrm{low}}$)
decreases with the increase of $\Delta f_{c}$ in the order $\mathcal{O}(1/(\Delta f_{c})^{2})$.
Besides, due to that $\overline{N}>\left|\gamma\right|,\forall\Delta\tau$,
the monotonicity between $\Delta f_{c}$ and $\textrm{CRB}_{\textrm{up}}$
($\textrm{CRB}_{\textrm{low}}$) is not influenced by $\Delta\tau$.

\begin{figure}[t]
\centering{}\includegraphics[width=6cm]{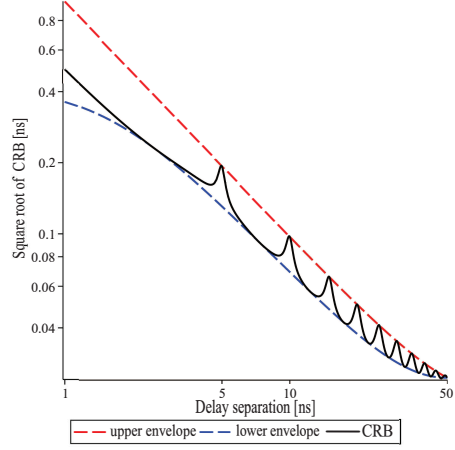}\caption{\label{fig:CRB_Closedform}An illustration of $C_{\Delta\tau}$ versus
delay separation.}
\end{figure}

\subsection{SRL for the Delay Resolution}

To further reveal the effects of frequency band apertures and phase
distortions on the fundamental limits of delay resolution and gain
more insights, we introduce the delay SRL \cite{SRL}, which is defined
as follows:
\begin{eqnarray}
 & \mathrm{SRL} & \triangleq\Delta\tau\nonumber \\
\text{\textrm{s.t.}} & \Delta\tau & =\sqrt{C_{\Delta\tau}}.\label{eq:P1}
\end{eqnarray}
 The delay SRL is the delay separation that is equal to its own root
squared CRB. In this definition, the delays can be exactly \textquotedblleft resolved\textquotedblright{}
when the standard deviation of the delay separation estimation is
equal to the true separation. It is difficult to obtain a closed-form
expression of delay SRL due to an intractable inverse operation corresponding
to the high-dimensional matrix $\mathbf{J}_{\boldsymbol{\eta}}$.
Therefore, we perform a numerical computation and provide useful insights
based on the numerical results.

\subsection{\label{subsec:SRL_simulation}Fundamental Limits Analysis Based on
Numerical Results}

\begin{figure}[t]
\centering{}\includegraphics[width=7.2cm]{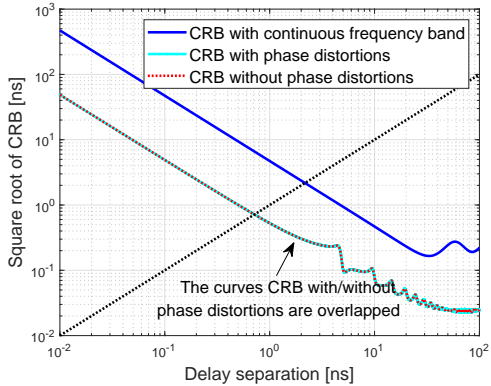}\caption{\label{fig:CRB)delayseparation}An illustration of the square root
CRB versus delay separation with $\left|\alpha_{1}\right|=1,\left|\alpha_{2}\right|=1$.}
\end{figure}
\begin{figure}[t]
\centering{}\includegraphics[width=7.2cm]{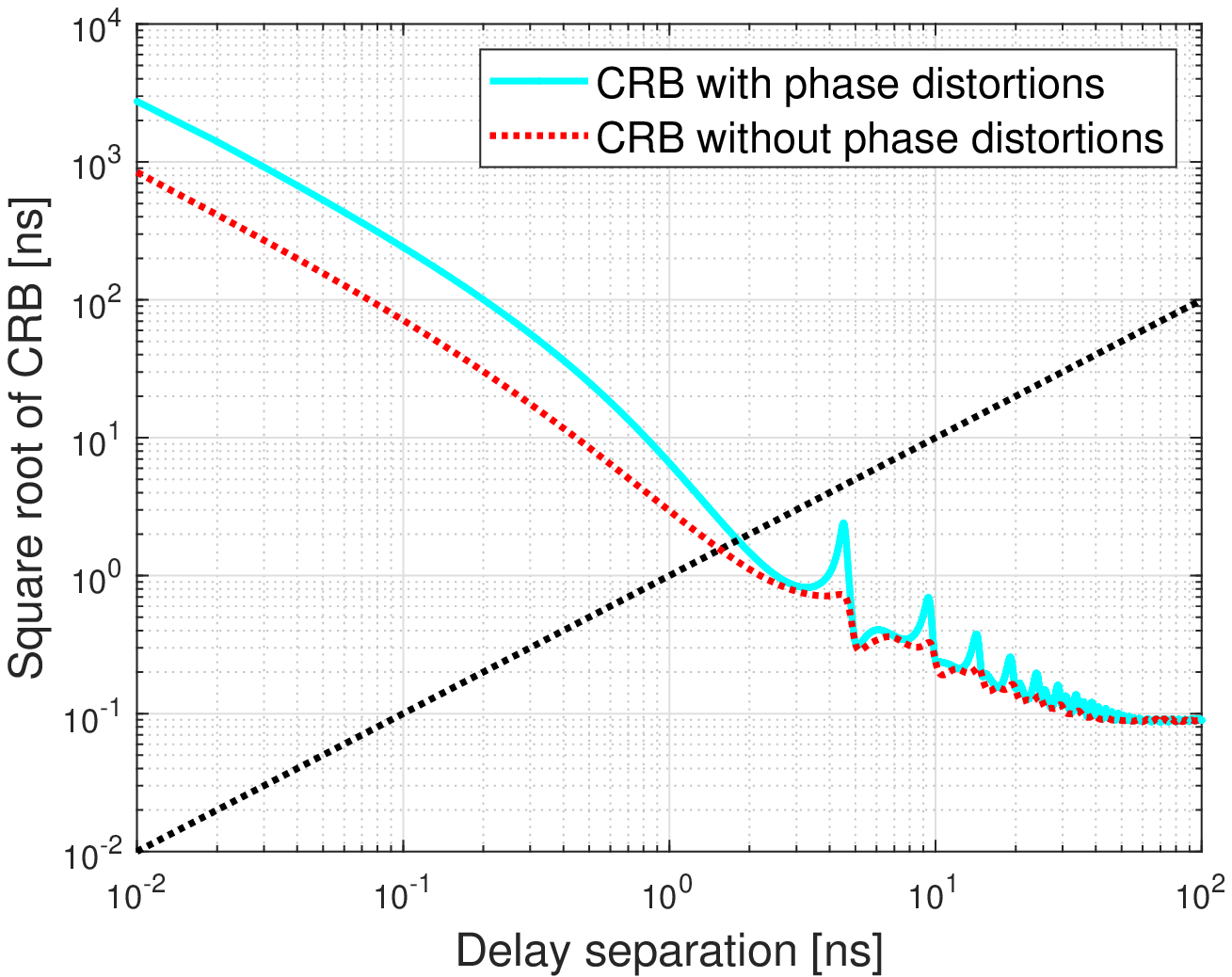}\caption{\label{fig:CRB)delayseparation-1}An illustration of the square root
CRB versus delay separation with $\left|\alpha_{1}\right|=1,\left|\alpha_{2}\right|=0.1$.}
\end{figure}
\begin{figure}[t]
\centering{}\includegraphics[width=7.2cm]{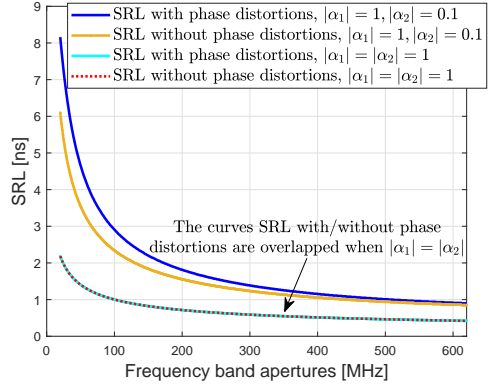}\caption{\label{fig:SRLvsFre}An illustration of the SRL versus frequency band
apertures.}
\end{figure}

In this subsection, we provide numerical results to study the effect
of frequency band apertures and phase distortions on the fundamental
limits of delay resolution. In the default setup, we consider that
the measurements are collected at $M=2$ subbands, with central frequencies
$f_{c,1}=1.8$ GHz and $f_{c,2}=2.0$ GHz, subcarrier spacing $f_{s,1}=f_{s,2}=78.125$
KHz, and subband bandwidth $B_{1}=B_{2}=20$ MHz. Besides, $\varphi_{m},\forall m$
and $\delta_{m},\forall m$ are generated following a uniform distribution
within $[0,2\pi]$ and a Gaussian distribution $\mathcal{N}\left(0,\sigma_{p}^{2}\right)$,
respectively. \textcolor{black}{The signal-to-noise ratio} (SNR) is
set as $15$ dB and the complex scalars are set as $\alpha_{1}=0.8+0.6j$
and $\alpha_{2}=0.6+0.8j$ with the unit amplitude.

Fig. \ref{fig:CRB)delayseparation} illustrates the square root of
the CRB for the delay separation versus the delay separation in different
scenarios, where the intersections of the CRB and the dotted black
line give the delay SRL. We compare the CRB with phase distortions
to two benchmarks: (i) An ideal scenario without phase distortions,
i.e., $\varphi_{m}=0,\forall m$ and $\delta_{m}=0,\forall m$; (ii)
An ideal scenario without phase distortions where only single contiguous
frequency band is employed with $40$ MHz equivalent bandwidth.

We have the following observations. First, the CRB based on two non-contiguous
frequency bands is lower than the CRB based on a single contiguous
frequency band with equivalent bandwidth, which leads to a higher
delay resolution (lower delay SRL) brought by the frequency band apertures.
Second, the curves of the CRB with/without phase distortions are completely
overlapped. Furthermore, we plot Fig. \ref{fig:CRB)delayseparation-1},
which illustrates the square root of CRB versus delay separation when
$\left|\alpha_{1}\right|=1,\left|\alpha_{2}\right|=0.1$. The curves
of the CRB with/without phase distortions are not overlapped anymore
and the phase distortions slightly decline the performance of delay
SRL. Therefore, a smaller difference between the amplitudes of the
multipath can significantly suppress the phase distortion interference.
In particular, when $\left|\alpha_{1}\right|=\left|\alpha_{2}\right|$,
the phase distortions have no effect on the delay SRL. \textcolor{blue}{To
justify this observation, we first derive the EFIM of $\mathbf{J}_{\boldsymbol{\eta}}$
for $\boldsymbol{\theta}$ given by
\begin{equation}
\left[\mathbf{J}_{\boldsymbol{\eta}}^{-1}\right]_{6\times6}\triangleq\mathbf{J}_{\mathrm{e}}^{-1}\left(\boldsymbol{\theta}\right)\overset{a}{=}(\mathbf{J}_{\boldsymbol{\theta}}-\mathbf{B}_{2}\mathbf{C}_{2}^{-1}\mathbf{B}_{2}^{T})^{-1}\overset{b}{=}\mathbf{J}_{\boldsymbol{\theta}}^{-1}+\boldsymbol{\varGamma},\label{eq:EFIM_3}
\end{equation}
where $\boldsymbol{\theta}=[\tau_{1},\tau_{2},\alpha_{R,1},\alpha_{R,2},\alpha_{I,1},\alpha_{I,2}]^{T}$
denotes the vector consisting of the unknown parameters in $\boldsymbol{\eta}$
except for phase distortions factors }\textcolor{blue}{\small{}$\varphi_{m}$
and $\delta_{m}$}\textcolor{blue}{, (\ref{eq:EFIM_3}-a) follows
the definition of EFIM, (\ref{eq:EFIM_3}-b) follows the Woodbury
identity \cite{Matrix_inverse}, $\mathbf{B}_{2}$ and $\mathbf{C}_{2}$
are the entries of $\mathbf{J}_{\boldsymbol{\eta}}\triangleq\left[\begin{array}{cc}
\mathbf{J}_{\boldsymbol{\theta}} & \mathbf{B}_{2}\\
\mathbf{B}_{2}^{T} & \mathbf{C}_{2}
\end{array}\right],$ and
\[
\boldsymbol{\varGamma}=\mathbf{J}_{\boldsymbol{\theta}}^{-1}\mathbf{B}_{2}(\mathbf{C}_{2}-\mathbf{B}_{2}^{T}\mathbf{J}_{\boldsymbol{\theta}}^{-1}\mathbf{B}_{2})^{-1}\mathbf{B}_{2}^{T}\mathbf{J}_{\boldsymbol{\theta}}^{-1}.
\]
Based on the symbolic computation, we find that $\boldsymbol{\varGamma}$
has a special symmetrical structure as $\boldsymbol{\varGamma}(1,1)=\boldsymbol{\varGamma}(1,2)=\boldsymbol{\varGamma}(2,1)=\boldsymbol{\varGamma}(2,2)$.
Then, $C_{\Delta\tau}$ derived from (\ref{eq:C_delattau}) can be
further equivalently transformed as
\begin{eqnarray}
C_{\Delta\tau} & = & \mathbf{J}_{\boldsymbol{\eta}}^{-1}(1,1)+\mathbf{J}_{\boldsymbol{\eta}}^{-1}(2,2)-\mathbf{J}_{\boldsymbol{\eta}}^{-1}(1,2)-\mathbf{J}_{\boldsymbol{\eta}}^{-1}(2,1)\nonumber \\
 & \overset{c}{=} & \mathbf{J}_{\boldsymbol{\theta}}^{-1}(1,1)+\mathbf{J}_{\boldsymbol{\theta}}^{-1}(2,2)-\mathbf{J}_{\boldsymbol{\theta}}^{-1}(1,2)-\mathbf{J}_{\boldsymbol{\theta}}^{-1}(2,1)\label{eq:C_bar2}\\
 &  & +\boldsymbol{\varGamma}(1,1)+\boldsymbol{\varGamma}(2,2)-\boldsymbol{\varGamma}(1,2)-\boldsymbol{\varGamma}(2,1)\nonumber \\
 & = & \mathbf{J}_{\boldsymbol{\theta}}^{-1}(1,1)+\mathbf{J}_{\boldsymbol{\theta}}^{-1}(2,2)-\mathbf{J}_{\boldsymbol{\theta}}^{-1}(1,2)-\mathbf{J}_{\boldsymbol{\theta}}^{-1}(2,1),\nonumber 
\end{eqnarray}
where (\ref{eq:C_bar2}-c) follows the equation (\ref{eq:EFIM_3}).
Finally, the CRB for the delay separation without $\varphi_{m}$ and
$\delta_{m}$ denoted as $\overline{C}_{\Delta\tau}$ is given by}

\textcolor{blue}{
\begin{eqnarray}
\overline{C}_{\Delta\tau} & = & \frac{\partial\Delta\tau}{\partial\boldsymbol{\theta}}^{T}\mathbf{J}_{\boldsymbol{\theta}}^{-1}\frac{\partial\Delta\tau}{\partial\boldsymbol{\theta}}\label{eq:C_delattau-1}\\
 & = & \mathbf{J}_{\boldsymbol{\theta}}^{-1}(1,1)+\mathbf{J}_{\boldsymbol{\theta}}^{-1}(2,2)-\mathbf{J}_{\boldsymbol{\theta}}^{-1}(1,2)-\mathbf{J}_{\boldsymbol{\theta}}^{-1}(2,1).\nonumber 
\end{eqnarray}
Therefore, from (\ref{eq:C_bar2}) and (\ref{eq:C_delattau-1}), we
have $\overline{C}_{\Delta\tau}=C_{\Delta\tau}$, which justified
the observations in Fig. \ref{fig:CRB)delayseparation}, i.e., the
curves SRL with/without phase distortions are overlapped.}

In Fig. \ref{fig:SRLvsFre}, we further investigate the effects of
frequency band apertures on the delay SRL. We fix $f_{c,1}$ and change
$f_{c,2}$ to generate different frequency band apertures. As can
be seen, the SRL decreases with the increase of frequency band apertures,
which is consistent with the monotonicity between $C_{\Delta\tau}$
and $\Delta f_{c}$ concluded in Subsection \ref{subsec:closed-form CRB}.
Besides, when $\left|\alpha_{1}\right|=\left|\alpha_{2}\right|$,
the SRL with/without phase distortions are always equal for different
frequency band apertures. In contrast, when $\left|\alpha_{1}\right|\ne\left|\alpha_{2}\right|$,
the SRL with/without phase distortions have a performance gap, that
declines as the frequency band apertures increase.

Finally, we summarize the key messages learned from the analysis in
this section.
\begin{enumerate}
\item \textbf{Monotonicity}: The CRB for delay separation $C_{\Delta\tau}$
decreases with the increase of frequency band apertures $\Delta f_{c}$
in the order $\mathcal{O}(1/(\Delta f_{c})^{2})$ and the monotonicity
is independent of delay separation $\Delta\tau$.
\item \textbf{Frequency band apertures gain}: (i) Under the equivalent total
bandwidth, the multiband sensing system with non-contiguous frequency
bands distribution reaps the extra frequency band apertures gain compared
to that with a single contiguous frequency band distribution, which
leads to a performance improvement of the delay resolution limit;
(ii) The SRL decreases with the increase of frequency band apertures.
\item \textbf{Phase distortions interference}: (i) The phase distortions
have relatively slight interference on delay SRL, which can be gradually
eliminated by increasing the frequency band apertures; (ii) A smaller
difference between the amplitudes of the multipath can suppress the
phase distortion interference better. Particularly, when $\left|\alpha_{1}\right|=\left|\alpha_{2}\right|$,
the delay SRL is not affected by phase distortions at all.
\end{enumerate}

\section{\label{sec:Fundamental Limit of Delay Estimation}Fundamental Limits
of Delay Estimation Error}

\subsection{CRB for the Delay Estimation Error}

we characterize the fundamental limits of delay estimation error by
a performance measure called delay error bound (DEB) as
\begin{equation}
\textrm{DEB}=\sqrt{\textrm{tr}\left\{ \left[\mathbf{J}_{\boldsymbol{\eta}}^{-1}\right]_{K\times K}\right\} },
\end{equation}
which is derived from the CRB. \textcolor{blue}{The notation $\textrm{tr}\left\{ \left[\mathbf{J}_{\boldsymbol{\eta}}^{-1}\right]_{K\times K}\right\} $
denotes that we first get the inverse of the $(3K$$+$$2M-1)$$\times$$(3K+2M-1)$
matrix $\mathbf{J}_{\boldsymbol{\eta}}$ and then select the first
$K\times K$ sub-matrix of $\mathbf{J}_{\boldsymbol{\eta}}^{-1}$
to take trace.} Then, we analyze the effect of frequency band apertures
and phase distortions on DEB based on numerical results. The parameters
are set as that in Subsection \ref{subsec:SRL_simulation} unless
otherwise specified.

\begin{figure*}[t]
\centering{}\negthinspace{}\negthinspace{}\subfloat[\label{fig:a}]{\centering{}\includegraphics[width=4.9cm]{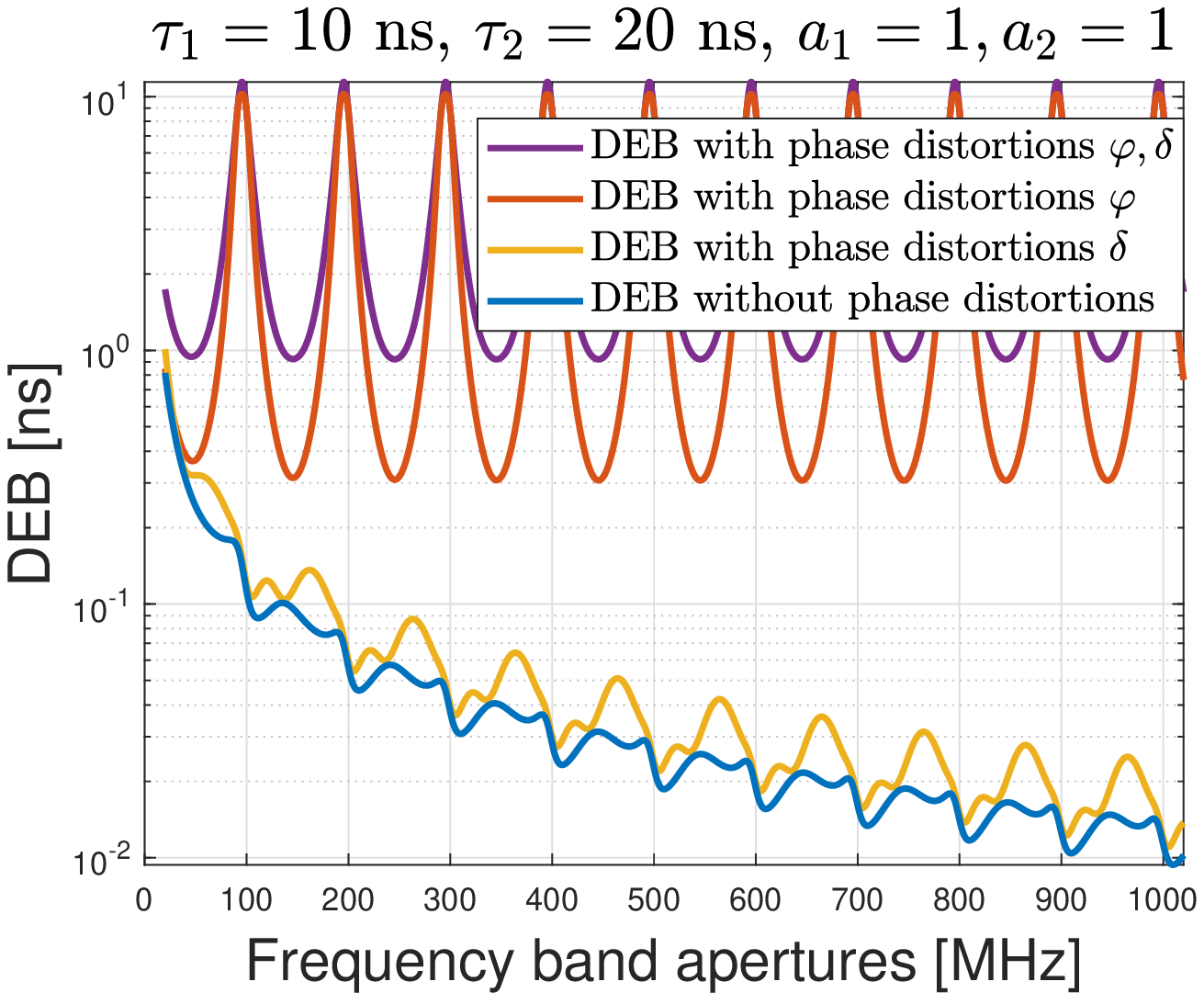}}\negthinspace{}\negthinspace{}\negthinspace{}\negthinspace{}\negthinspace{}\negthinspace{}\negthinspace{}\subfloat[\label{fig:b}]{\centering{}\includegraphics[width=4.9cm]{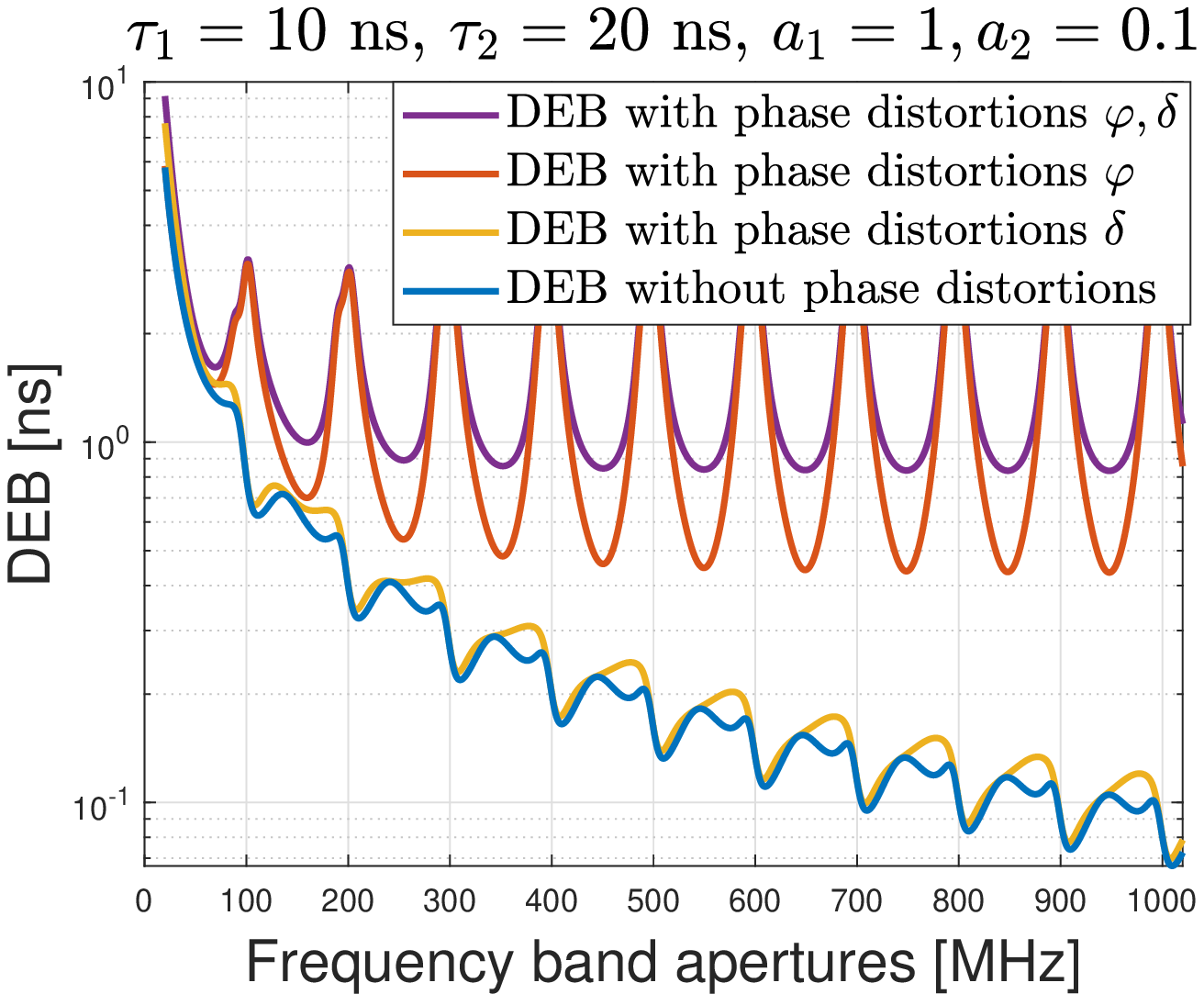}}\negthinspace{}\negthinspace{}\subfloat[\label{fig:c}]{\centering{}\includegraphics[width=4.9cm]{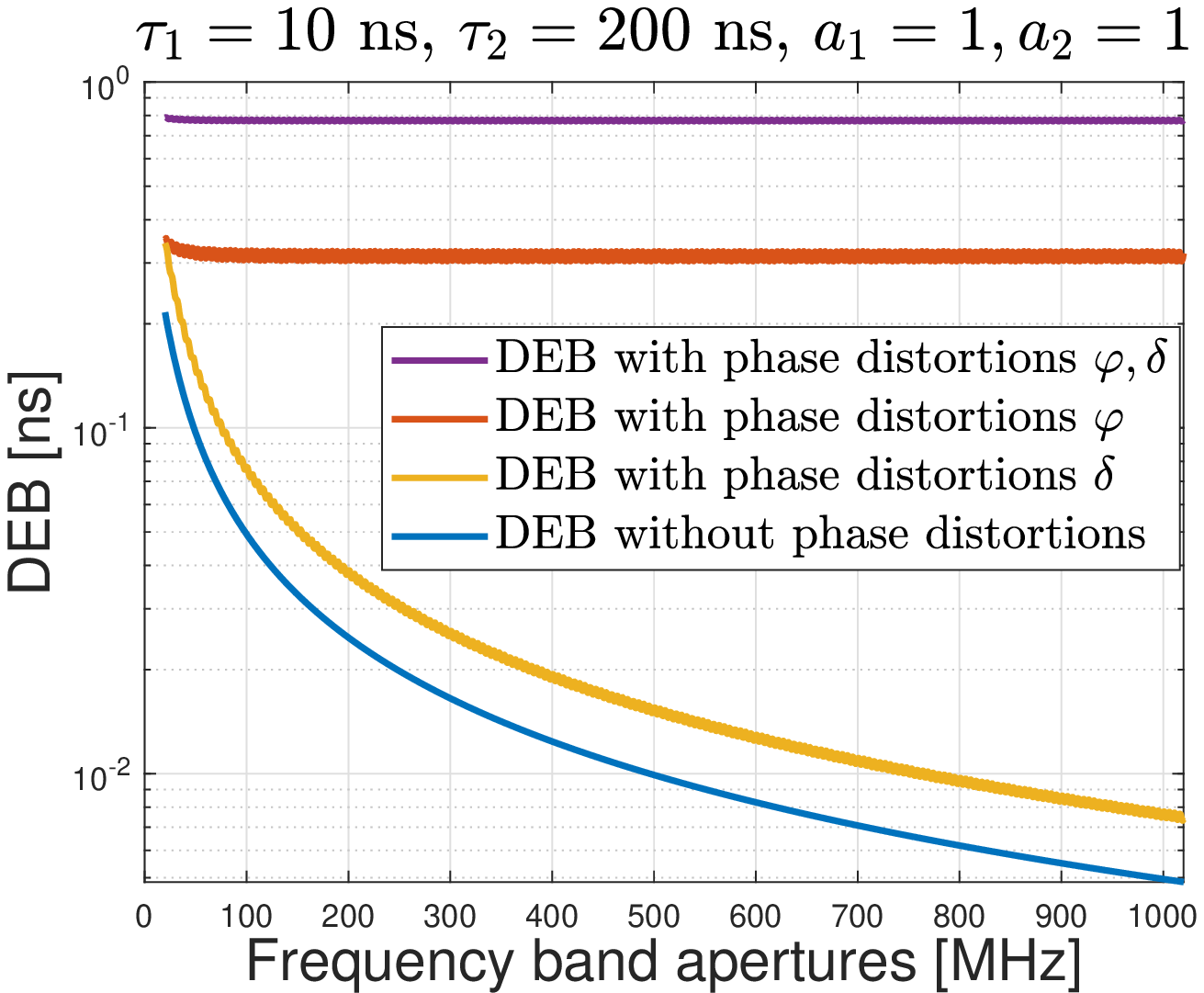}}\negthinspace{}\negthinspace{}\negthinspace{}\negthinspace{}\negthinspace{}\negthinspace{}\negthinspace{}\subfloat[\label{fig:d}]{\centering{}\includegraphics[width=4.9cm]{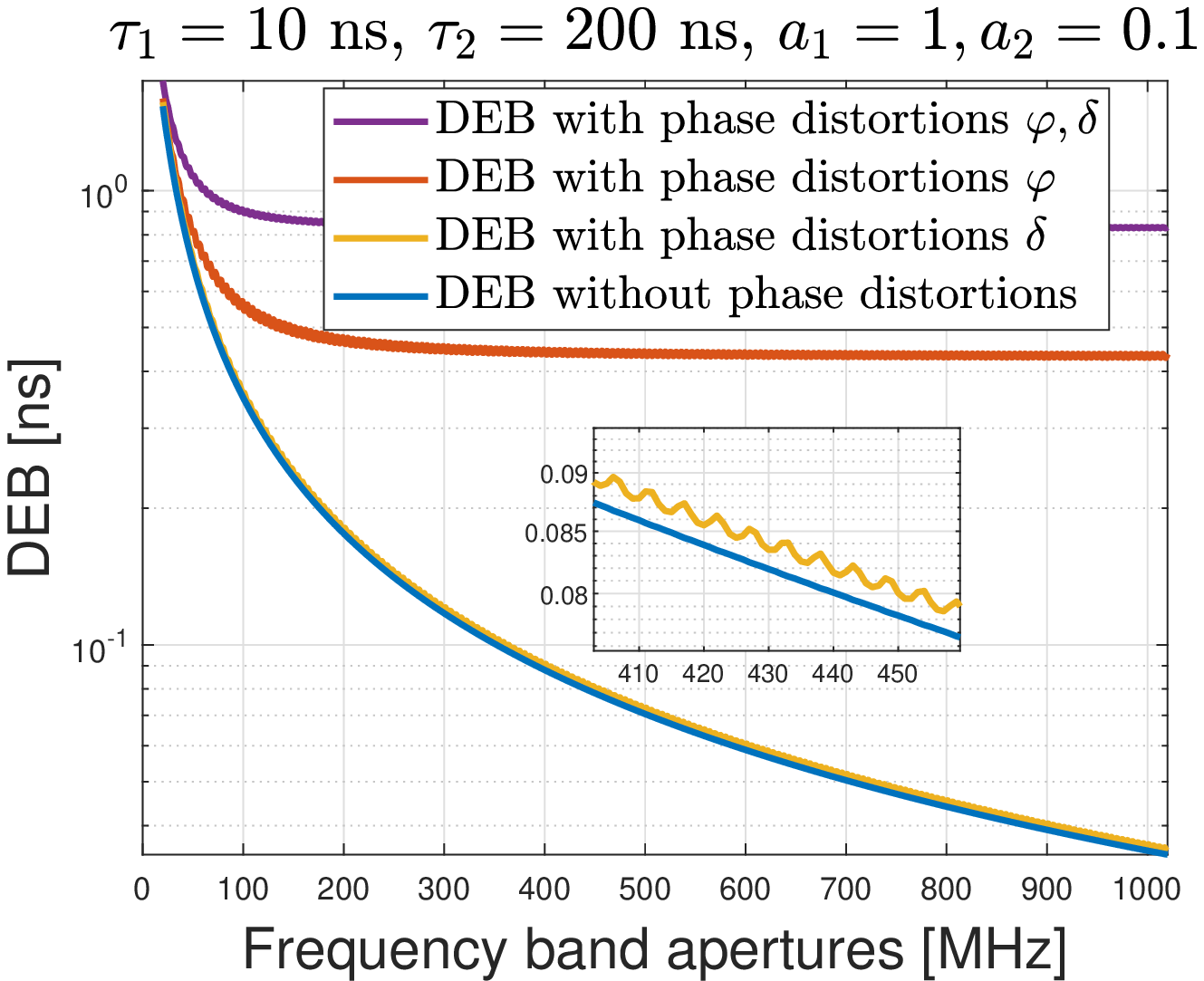}}\caption{\label{fig:CRB1}An illustration of DEB versus frequency band apertures.}
\end{figure*}

In Fig. \ref{fig:CRB1}, we consider four scenarios: (1) An ideal
scenario without phase distortions; (2) A scenario with only random
phase offset $\boldsymbol{\varphi}$; (3) A scenario with only receiver
timing offset $\boldsymbol{\delta}$; (4) A scenario with both phase
distortions factors $\boldsymbol{\varphi}$ and $\boldsymbol{\delta}$.
For each scenario, we consider different values of $\tau_{2}$ and
$a_{2}$, as shown in Fig. \ref{fig:a}-Fig. \ref{fig:d}.

It is observed that both $\boldsymbol{\varphi}$ and $\boldsymbol{\delta}$
lead to a larger DEB. Besides, in Fig. \ref{fig:a}, the DEB without
phase distortions decreases as the frequency band aperture increases
with a slight fluctuation. The fluctuation can be justified by the
trigonometric term in (\ref{eq:FIM_compact}) with regard to $f_{c,m}$,
e.g., $\cos\left(2\pi f_{c,m}\Delta\tau+\phi_{1}\right)$, which is
a periodic function with period $1/\Delta\tau$. In fact, we can observe
that a harmonic component exists in all scenarios with period $1/\Delta\tau$
and the scenarios with $\boldsymbol{\varphi}$ have the most violent
fluctuation of DEB. It reveals the difficulty of exploiting frequency
band apertures gain for any algorithms in the presence of $\boldsymbol{\varphi}$.
However, as shown in Fig. \ref{fig:b}-Fig. \ref{fig:d}, if we increase
$\Delta\tau$ and the difference between the amplitudes $a_{1}$ and
$a_{2}$, we will observe that the DEB decreases as the frequency
band apertures increase with much weaker fluctuation, indicating that
the frequency band apertures contribute to improving the delay estimation
accuracy though in the presence of phase distortions.

\begin{figure*}[tbh]
\begin{centering}
\textcolor{black}{}%
\begin{minipage}[t]{0.45\textwidth}%
\begin{center}
\subfloat[\label{fig:DEB)delta_a}DEB in the presence of phase distortion factors
$\boldsymbol{\varphi}$ and $\boldsymbol{\delta}$.]{\begin{centering}
\includegraphics[width=72mm]{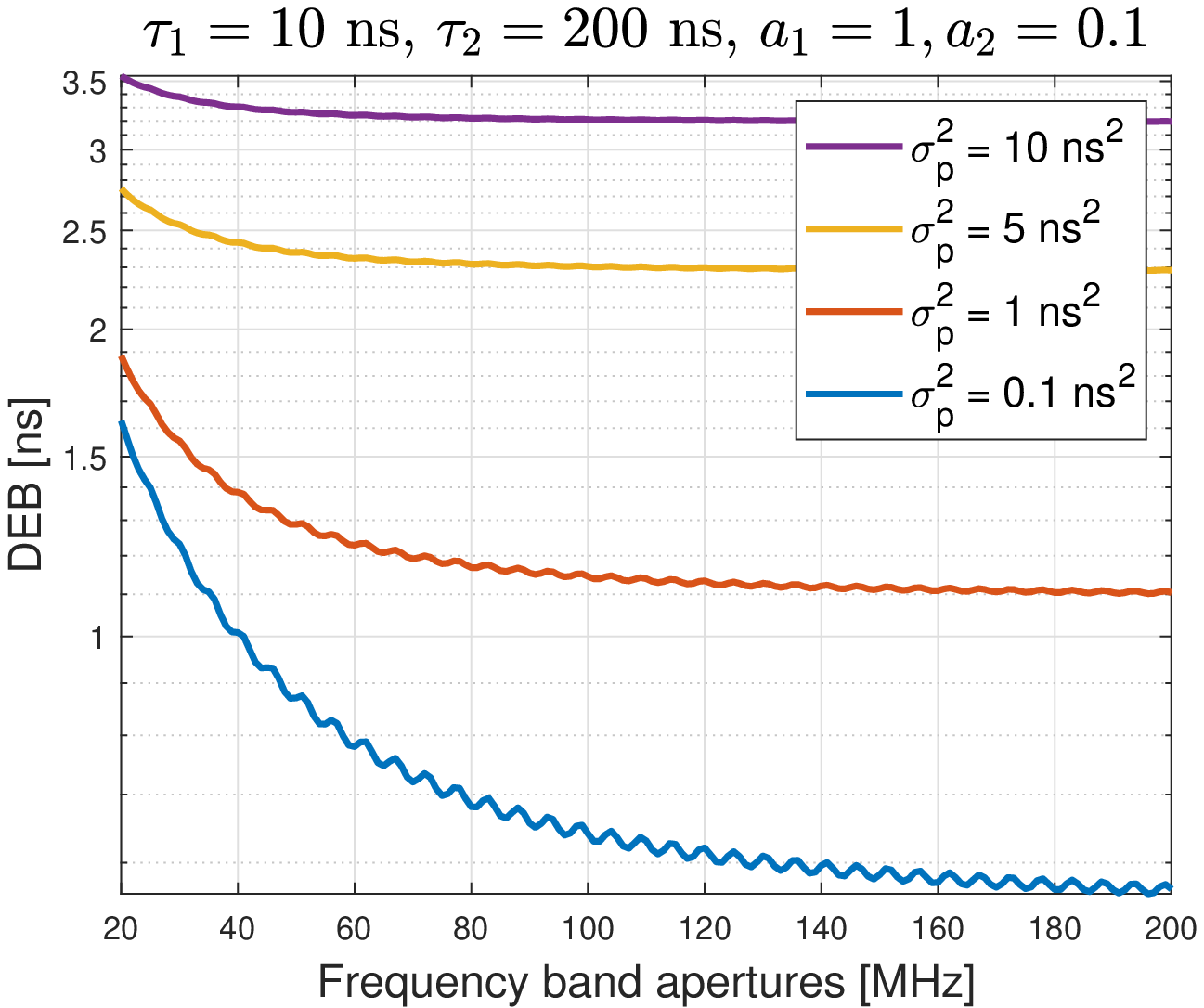}
\par\end{centering}
}
\par\end{center}%
\end{minipage}\textcolor{black}{\hfill{}}%
\begin{minipage}[t]{0.45\textwidth}%
\begin{center}
\subfloat[\label{fig:DEB)delta_b}DEB in the presence of phase distortion factor
$\boldsymbol{\delta}$.]{\begin{centering}
\includegraphics[width=72mm]{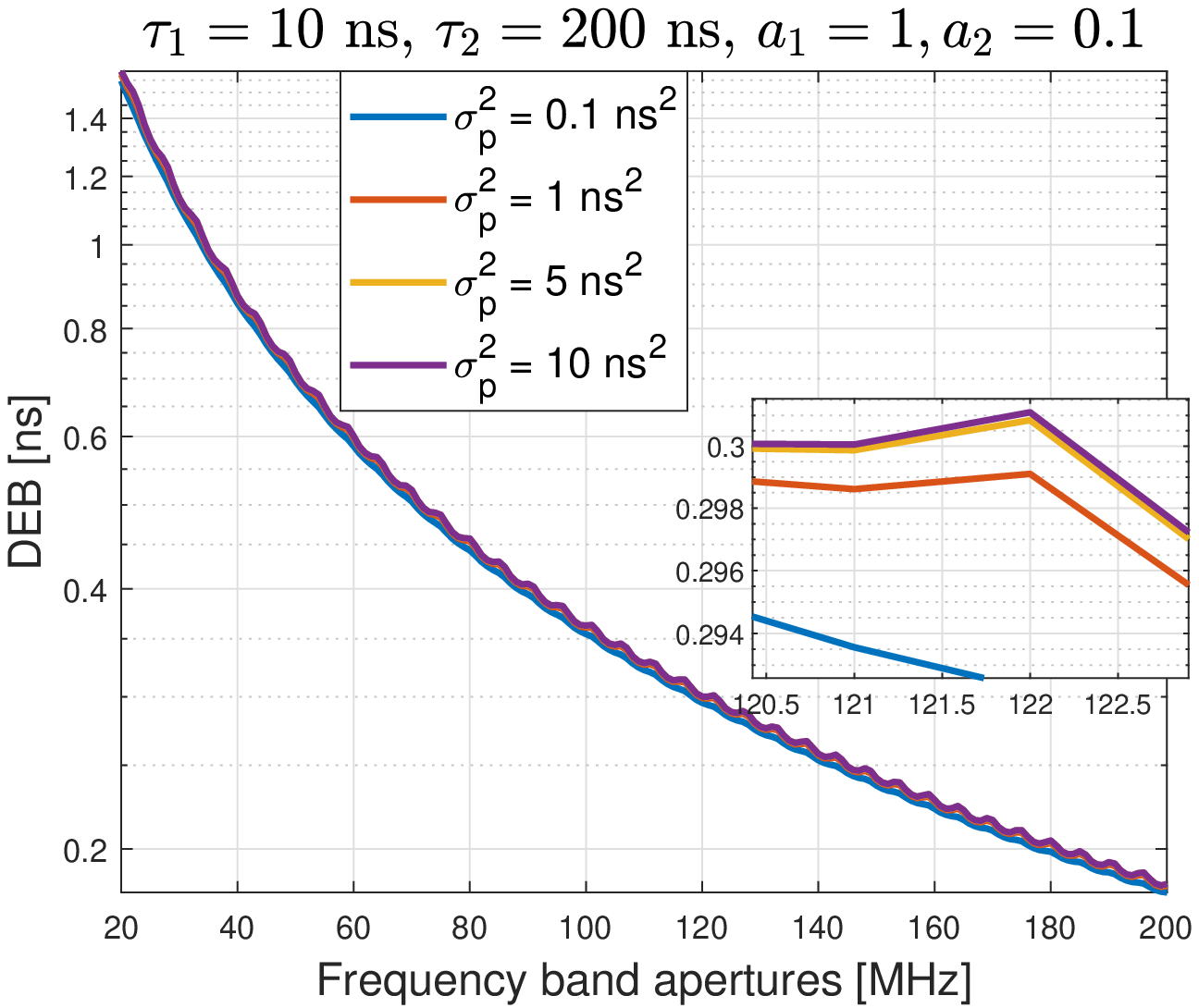}
\par\end{centering}
}
\par\end{center}%
\end{minipage}
\par\end{centering}
\caption{\label{fig:CRB_delta}An illustration of DEB versus frequency band
apertures for different $\sigma_{p}^{2}$.}
\end{figure*}

In Fig. \ref{fig:CRB_delta}, we further investigate the effect of
$\boldsymbol{\delta}$ on DEB. As can be seen, the DEB increases with
$\sigma_{p}$, which is reasonable since larger $\sigma_{p}$ leads
to little prior information of $\boldsymbol{\delta}$. Furthermore,
in Fig. \ref{fig:DEB)delta_a}, the value of $\sigma_{p}$ dominates
the behavior of DEB while in Fig. \ref{fig:DEB)delta_b}, increasing
the value of $\sigma_{p}$ incurs only negligible performance loss.
It is because the prior information of $\boldsymbol{\delta}$ is helpful
to eliminate the ambiguity of signal model (\ref{eq:refined_signal})
in the presence of $\boldsymbol{\varphi}$ and $\boldsymbol{\delta}$.
When in the absence of $\boldsymbol{\varphi}$, the signal model (\ref{eq:refined_signal})
does not exist ambiguity anymore and thus the effect of prior information
becomes negligible.

\subsection{ZZB for the Delay Estimation Error}

\textcolor{blue}{CRB is a local bound that may provide inaccurate
predictions under conditions of low SNR or large frequency band apertures.
Therefore, providing a global bound for the delay estimation error
is necessary, which is capable of providing more accurate predictions
of the performance of estimators over the full range of SNR and frequency
band apertures.}

\textcolor{blue}{In this subsection, we adopt a widely employed global
bound ZZB as the fundamental limits of the multiband delay estimation
problem. It is also a Bayesian bound that incorporates prior information
of the unknown parameters and is not limited to unbiased estimates
\cite{ZZB1,ZZB2,ZZB3,ZZB4}.} We consider a scenario without phase
distortions, where $K=1$ with unknown parameters $\boldsymbol{\eta}_{1}=\left[\tau_{1},\phi_{1}\right]^{T}$
for simplicity, since the calculation of ZZB for $K\geq2$ is extremely
difficult. The signal model in (\ref{eq:refined_signal}) can be reformulated
as
\begin{equation}
y_{m,n}=\left|\alpha_{1}\right|e^{j\phi_{1}}e^{-j2\pi\left(f_{c,m}^{\prime}+nf_{s,m}\right)\tau_{1}}s_{m,n}+w_{m,n}.\label{eq:ZZB_singlepath}
\end{equation}

\textcolor{blue}{The development of the ZZB links the delay estimation
problem to a hypothesis testing problem that discriminates a signal
at two possible delays. For a received signal at one of the two possible
parameters value $\boldsymbol{\beta}$ or $\boldsymbol{\beta+\mathbf{e}}$,
where $\mathbf{e}=\left[e_{\tau_{1}},e_{\phi_{1}}\right]^{T}$ denotes
an offset, the hypothesis test denoted as $\mathcal{P}_{H}$ is given
by}
\[
\begin{aligned} & \textrm{Decide}\thinspace\mathrm{\mathcal{H}}_{0}\negthinspace:\negthinspace\boldsymbol{\eta}_{1}\negthinspace=\negthinspace\boldsymbol{\beta}\text{ if }\boldsymbol{u}^{T}\negthinspace\hat{\boldsymbol{\eta}}_{1}\negthinspace\leq\negthinspace\boldsymbol{u}^{T}\negthinspace\boldsymbol{\beta}\negthinspace+\negthinspace\frac{h}{2};\mathbf{y}\sim p_{\mathbf{y}|\boldsymbol{\eta}_{1}}\negthinspace(\mathbf{y}|\boldsymbol{\beta}),\\
 & \textrm{Decide}\thinspace\mathrm{\mathcal{H}}_{1}\negthinspace:\negthinspace\boldsymbol{\eta}_{1}\negthinspace=\negthinspace\boldsymbol{\beta}\negthinspace+\negthinspace\mathbf{e}\text{ if }\negthinspace\boldsymbol{u}^{T}\negthinspace\hat{\boldsymbol{\eta}}_{1}\negthinspace\negthinspace>\negthinspace\boldsymbol{u}^{T}\negthinspace\boldsymbol{\beta}\negthinspace+\negthinspace\frac{h}{2};\mathbf{y}\negthinspace\sim\negthinspace p_{\mathbf{y}|\boldsymbol{\eta}_{1}}\negthinspace(\mathbf{y}|\boldsymbol{\beta}\negthinspace+\negthinspace\mathbf{e}),
\end{aligned}
\]
with prior probabilities $\textrm{Pr}(\mathcal{H}_{0})=P_{0}$ and
$\textrm{Pr}(\mathcal{H}_{1})=P_{1}$. \textcolor{blue}{A more detailed
discussion of the hypothesis test formulation may be found in \cite{ZZB3}}.
Note that $\boldsymbol{u}$ can be any 2-dimensional vector and $\hat{\boldsymbol{\eta}}_{1}$
denotes an estimator of $\boldsymbol{\eta}_{1}$. Let $P_{\min}(\boldsymbol{\beta},\boldsymbol{\beta}+\mathbf{e})$
denotes minimal probability of error achieved by the optimum detection
scheme in making the above decision. Then, the ZZB for the quadratic
form of the MSE matrix is given by \cite{ZZB2}
\begin{equation}
\begin{aligned}\mathbf{u}^{T}\boldsymbol{\Phi}\mathbf{u}\geq & \frac{1}{2}\int_{0}^{\infty}\mathcal{V}\{\max_{\mathbf{e}:\mathbf{u}^{T}\mathbf{e}=h}\int_{\Theta}\left[p_{\boldsymbol{\eta}_{1}}(\boldsymbol{\beta})+p_{\boldsymbol{\eta}_{1}}(\boldsymbol{\beta}+\mathbf{e})\right]\\
 & \times P_{\min}(\boldsymbol{\beta},\boldsymbol{\beta}+\mathbf{e})d\boldsymbol{\beta}\}hdh,
\end{aligned}
\label{eq:ZZB_ori}
\end{equation}
where $\boldsymbol{\Phi}\triangleq\mathbb{E}_{\mathbf{y},\boldsymbol{\eta}_{1}}\left\{ (\hat{\boldsymbol{\eta}}_{1}-\boldsymbol{\eta}_{1})(\hat{\boldsymbol{\eta}}_{1}-\boldsymbol{\eta}_{1})^{H}\right\} $
denotes the MSE matrix, $\Theta$ denotes the region in which $\boldsymbol{\eta}_{1}$
is defined, $\mathcal{V}\{\cdot\}$ denotes the valley-filling function
\cite{ZZB3}, and $p_{\boldsymbol{\eta}_{1}}(\cdot)$ denotes the
prior distribution of the unknown parameters vector $\boldsymbol{\eta}_{1}$.
Given the prior distribution, the ZZB is evaluated involving an integral
of a product for the known prior distribution and the minimum detection
error probability. Therefore, to compute the ZZB, the minimum detection
error probability is the major unknown component needed to calculate.
In the subsequent content, we will first evaluate $P_{\min}(\boldsymbol{\beta},\boldsymbol{\beta}+\mathbf{e})$
by calculating the error probability of the optimum log-likelihood
ratio (LLR) test for $\mathcal{P}_{H}$, and then substitute the expression
of $P_{\min}(\boldsymbol{\beta},\boldsymbol{\beta}+\mathbf{e})$ into
(\ref{eq:ZZB_ori}) to compute the ZZB.

Consider a pair of equally likely hypotheses in $\mathcal{P}_{H}$,
where the prior probability $\textrm{Pr}\left(\mathcal{H}_{0}\right)=\textrm{Pr}\left(\mathcal{H}_{1}\right)=1/2$.
Then the minimum probability of error $P_{\min}(\boldsymbol{\beta},\boldsymbol{\beta}+\mathbf{e})$
can be obtained from the LLR test \cite{LLR} as
\[
P_{\min}(\boldsymbol{\beta},\boldsymbol{\beta}+\mathbf{e})=\frac{1}{2}\textrm{Pr}\left(\zeta<0\mid H_{0}\right)+\frac{1}{2}\textrm{Pr}\left(\zeta>0\mid H_{1}\right),
\]
where $\zeta$ is the LLR for the hypothesis test given by
\[
\zeta=\ln p_{\mathbf{y}\mid\boldsymbol{\eta}_{1}}(\mathbf{y}\mid\boldsymbol{\beta})-\ln p_{\mathbf{y}\mid\boldsymbol{\eta}_{1}}(\mathbf{y}\mid\boldsymbol{\beta}+\mathbf{e}).
\]
Let $\mathbf{y}_{i}\triangleq\mathbf{y}|\mathcal{H}_{i}$, $\zeta_{i}\triangleq\zeta|\mathcal{H}_{i},i=0,1$,
then $\zeta_{i}\propto\textrm{Re}\{(\boldsymbol{u}_{0}-\boldsymbol{u}_{1})^{H}\mathbf{y}_{i}\}$,
$\forall i$, where
\[
\boldsymbol{u}_{0}=[u_{0}(1,-\frac{N_{m}-1}{2}),...,u_{0}(M,\frac{N_{m}-1}{2})]^{T},
\]
whose $(m,n)$-th element is given by
\[
\begin{aligned}u_{0}(m,n) & =\left|\alpha_{1}\right|e^{j\beta_{\phi_{1}}}e^{-j2\pi(f_{c,m}^{\prime}+nf_{s,m})\beta_{\tau_{1}}},\forall m,n.\end{aligned}
\]
Similarly, $\boldsymbol{u}_{1}$ is a vector whose $(m,n)$-th element
is
\[
u_{1}(m,n)=\left|\alpha_{1}\right|e^{j(\beta_{\phi_{1}}+e_{\phi_{1}})}e^{-j2\pi(f_{c,m}^{\prime}+nf_{s,m})(\beta_{\tau_{1}}+e_{\tau_{1}})}.
\]
Note that the LLR $\zeta$ is a linear combination of Gaussian variables,
we can get the expectation and variance of $\zeta_{0}$ and $\zeta_{1}$
as
\[
\begin{aligned}\mathbb{E}\left[\zeta_{0}\right] & =-\mathbb{E}\left[\zeta_{1}\right]\propto\left|\alpha_{1}\right|^{2}(N\negthinspace-\negthinspace\sum_{m,n}\cos(-2\pi f_{m,n}e_{\tau_{1}}\negthinspace+\negthinspace e_{\phi_{1}})),\\
\mathbb{D}\left[\zeta_{0}\right] & =\mathbb{D}\left[\zeta_{1}\right]\propto\sigma_{ns}^{2}\left|\alpha_{1}\right|^{2}(N\negthinspace-\negthinspace\sum_{m,n}\cos(-2\pi f_{m,n}e_{\tau_{1}}\negthinspace+\negthinspace e_{\phi_{1}})).
\end{aligned}
\]
As can be seen, $P_{\min}(\boldsymbol{\beta},\boldsymbol{\beta}+\mathbf{e})$
is only a function of the offset $\mathbf{e}$ and thus can be denoted
by $P_{\min}(\mathbf{e})$, which is given by
\[
\begin{aligned}P_{\min}(\mathbf{e}) & =Q\left(\frac{\mathbb{E}\left[\zeta_{0}\right]}{\sqrt{\mathbb{D}\left[\zeta_{0}\right]}}\right)\\
 & =Q\left(\frac{\left|\alpha_{1}\right|}{\sqrt{\sigma_{ns}^{2}}}\sqrt{N\negthinspace-\negthinspace\sum_{m,n}\cos\left(-2\pi f_{m,n}e_{\tau_{1}}+e_{\phi_{1}}\right)}\right),
\end{aligned}
\]
where $Q\{\cdot\}$ denotes the tail distribution function of the
standard normal distribution, i.e., $Q(x)=\frac{1}{\sqrt{2\pi}}\int_{x}^{\infty}e^{-\frac{v^{2}}{2}}dv$.
Assume that the unknown parameters $\tau_{1}$ and $\phi_{1}$ are
independent random variables, which are uniformly distributed in $[0,D]$
\textcolor{blue}{ns} and $[0,2\pi]$, respectively. Then, substituting
$\mathbf{u}=[1,0]^{T}$ in (\ref{eq:ZZB_ori}), the ZZB for the delay
estimation error is given by
\[
\begin{aligned}\textrm{ZZB}_{\tau_{1}} & =\left.\boldsymbol{u}^{T}\boldsymbol{\Phi}\boldsymbol{u}\right|_{u=[1,0]^{T}}=\boldsymbol{\Phi}(1,1)\\
 & \geq\frac{1}{2\pi D}\int_{0}^{D}\negthinspace e_{\tau_{1}}\negthinspace\mathcal{V}\left\{ \left(D\negthinspace-\negthinspace e_{\tau_{1}}\right)\max_{e_{\phi_{1}}}\left(2\pi\negthinspace-\negthinspace e_{\phi_{1}}\right)P_{\min}(\mathbf{e})\right\} de_{\tau_{1}}.
\end{aligned}
\]

There is no closed-form expression of $\textrm{ZZB}_{\tau_{1}}$,
so we perform a numerical computation to obtain its value. We compare
the derived ZZB with MSEs of the MAP delay estimator and the expected
CRB (ECRB) \cite{ECRB}. \textcolor{blue}{The MAP estimation results
are obtained based on a 2-dimensional exhausted search for variables
$\tau_{1}$ and $\phi_{1}$. The root MSE (RMSE) of MAP estimates
is then calculated over 200 Monte Carlo trials.} The ECRB is obtained
by taking the expectation of the conditional CRB with respect to the
random but unknown parameters $\boldsymbol{\eta}_{1}$, namely
\begin{equation}
\mathrm{ECRB}=\mathbb{E}_{\boldsymbol{\eta}_{1}}\left[\mathrm{CRB}(\boldsymbol{\eta}_{1})\right].
\end{equation}

\begin{figure}[t]
\centering{}\includegraphics[width=7.2cm]{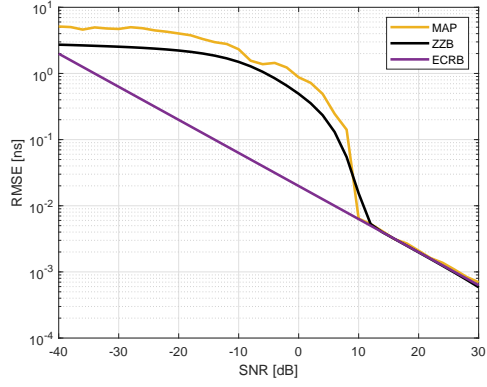}\textcolor{blue}{\caption{\textcolor{blue}{\label{fig:ZZB_SNR_MAP}An illustration of ZZB with
MAP and ECRB comparison versus SNR.}}
}
\end{figure}

In the default setup, we consider two subbands with subcarrier spacing
$f_{s,1}=f_{s,2}=78.125$ KHz and bandwidth $B_{1}=B_{2}=B$, where
$B=20$ MHz. SNR is $10$ dB and $D=10$. From Fig. \ref{fig:ZZB_SNR_MAP}
where the frequency band aperture $\Delta f_{c}=0.5$ GHz, the curve
of ZZB versus SNR can be divided into three regions. In the low SNR
region, the ZZB provides a tighter bound than ECRB. Besides, the ZZB
reaches a plateau equal to the standard deviation of the prior distribution
of $\tau_{1}$, that can be computed as $\sqrt{D^{2}/12}$, due to
that the sensing performance is mainly dominated by prior information
when SNR is low. For the high SNR region, the ZZB merges with the
ECRB and MAP. Besides, the ZZB predicts the MAP threshold behavior
and a transition region is observed between the low and high SNR regions.

\begin{figure}[t]
\centering{}\includegraphics[width=7.2cm]{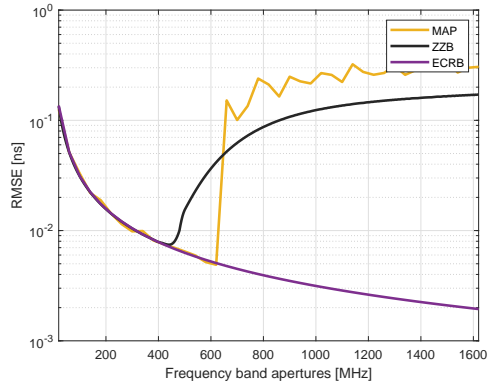}\textcolor{blue}{\caption{\textcolor{blue}{\label{fig:ZZBFre}An illustration of ZZB with MAP
and ECRB comparison versus frequency band apertures.}}
}
\end{figure}

Fig. \ref{fig:ZZBFre} displays the ZZB, ECRB, and MSEs of the MAP
estimates as a function of frequency band apertures. As can be seen,
the ZZB provides a bound at least as tight or tighter than the ECRB
in most frequency band apertures regions. Besides, the MAP threshold
behavior emerges as the frequency band apertures increase, i.e., the
RMSE of MAP estimator decreases first and then rapidly increases with
the increase of frequency band apertures. It can be justified that
though larger frequency band apertures result in a sharper mainlobe
of the likelihood function, which reveals a potential sensing performance
gain, it also leads to a multimodal likelihood function that has more
sidelobes. The MAP estimator will be affected by ambiguities created
by the sidelobes. Moreover, similar to the observations in Fig. \ref{fig:ZZB_SNR_MAP},
the ZZB again predicts the MAP threshold behavior while the ECRB does
not track the MAP threshold behavior. It is reasonable since the ECRB
is a local bound whereas the ZZB is a global bound. Inspired by \textcolor{blue}{these}
observations, the frequency band apertures should be restricted to
a limited range in practical multiband sensing systems, in order to
avoid causing a performance loss and fully exploit the frequency band
apertures gain.


\begin{figure}[t]
\centering{}\includegraphics[width=7.2cm]{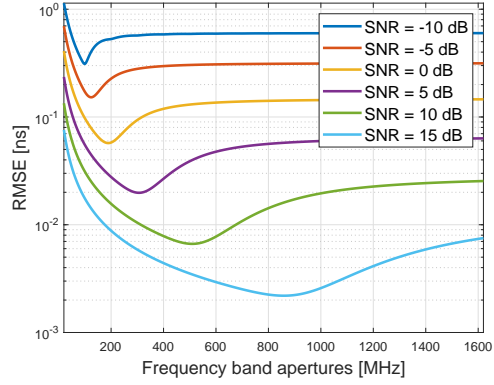}\caption{\label{fig:ZZB_SNR}An illustration of ZZB versus frequency band apertures
for different SNRs.}
\end{figure}

\begin{figure}[t]
\centering{}\includegraphics[width=7.2cm]{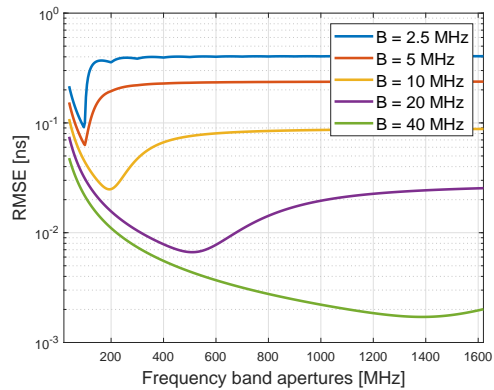}\caption{\label{fig:ZZB_B}An illustration of ZZB versus frequency band apertures
for different bandwidths.}
\end{figure}

Fig. \ref{fig:ZZB_SNR} and Fig. \ref{fig:ZZB_B} illustrate the effect
of SNR and bandwidths on the ZZB, respectively. It can be seen that
the ZZB decreases with the increase of SNR or bandwidth. Besides,
the threshold behavior emerges in a larger frequency band aperture
as the SNR or bandwidth increases, due to that the ambiguities caused
by sidelobes are significantly reduced.

\begin{figure}[t]
\centering{}\includegraphics[width=7.2cm]{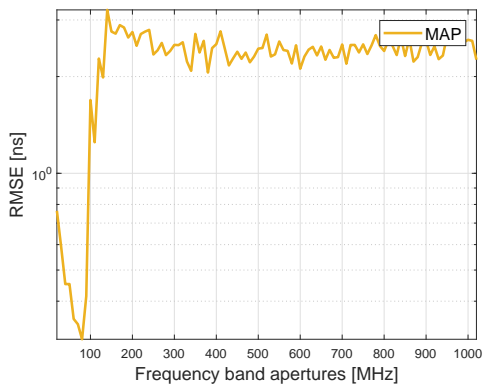}\caption{\label{fig:ZZB_twopath}An illustration of MAP estimator versus frequency
band aperture based on a multiple targets signal model.}
\end{figure}

Though the above results are obtained based on a single target signal
model, the observations can also be observed based on a multiple targets
signal model with phase distortions considered. Due to the difficulty
of computing the ZZB associated to a multi-parameter estimation problem,
we plot Fig. \ref{fig:ZZB_twopath} to just illustrate the RMSE of
the MAP estimator as a function of frequency band apertures based
on the multiple targets signal model (\ref{eq:refined_signal}), where
the parameters are set as that in Subsection \ref{subsec:SRL_simulation}.
As can be seen, the MAP threshold behavior appears as expected.

Finally, we summarize the key messages learned from the analysis in
this section.
\begin{enumerate}
\item \textbf{Monotonicity}: (i) The DEB without phase distortions factors
$\boldsymbol{\varphi}$ decreases as the frequency band aperture increases
with a slight fluctuation; (ii) Generally, both the DEB with/without
phase distortions fluctuate as a function of frequency band apertures
due to the existence of trigonometric terms whose period is $1/\Delta\tau$.
\item \textbf{Interference of} \textbf{random phase offset} $\boldsymbol{\varphi}$:
The existence of $\boldsymbol{\varphi}$ leads to a larger DEB with
violent fluctuation and thus makes it difficult to exploit the frequency
band apertures gain for any methods. However, when the targets are
distinguishable with significantly different time delay and amplitudes,
the DEB decreases as the frequency band apertures increase smoothly,
which unveils a delay estimation performance gain brought by the frequency
band apertures even though in the presence of phase distortions.
\item \textbf{Interference of receiver timing offset} $\boldsymbol{\delta}$:
The existence of $\boldsymbol{\delta}$ leads to a larger DEB, but
this negative effect can be suppressed by increasing the prior information
of $\boldsymbol{\delta}$. Specifically, for the signal model (\ref{eq:refined_signal})
in the presence of $\boldsymbol{\varphi}$ and $\boldsymbol{\delta}$,
the DEB significantly decreases with $\sigma_{p}$ since the prior
information eliminates the signal model ambiguity efficiently. For
the signal model (\ref{eq:refined_signal}) in the absence of $\boldsymbol{\varphi}$,
the effect of prior information becomes relatively negligible since
the signal model does not exist ambiguity anymore.
\item \textbf{ZZB behavior}: (i) The ZZB provides a tighter bound than the
ECRB in all frequency band apertures regions and SNR regions; (ii)
The ZZB predicts the MAP threshold behavior, which emerges as the
frequency band apertures increase or the SNR decreases. Hence, the
subbands need to be selected carefully and the frequency band apertures
should be restricted to a limited range in practical multiband sensing
systems, in order to fully exploit the frequency band apertures gain;
(iii) Increasing the the SNR or bandwidth can delay the occurrence
of the threshold behavior as frequency band apertures increase.
\end{enumerate}

\section{\label{sec:The Multiband Optimization Problem}Optimization of Multiband
Sensing Systems}

\textcolor{blue}{The fundamental limits analysis discussed above implies
that it is possible to improve the sensing performance of the multiband
sensing system by system parameter optimization, e.g., increasing
the frequency band apertures of the subbands may improve the system's
sensing resolution.}

\textcolor{blue}{Therefore, in this section, we present the optimization
of multiband sensing systems with the objective of minimizing the
fundamental limit, delay SRL under a few practical constraints. The
reason we adopt the delay SRL as the objective function is that the
delay SRL is less affected by the phase distortions than DEB and requires
less information to calculate (e.g., only estimated mean amplitude
information is required to calculate the SRL in practical scenarios,
as will be explained later). In fact, the optimized results with the
objective of minimizing the delay SRL is also effective to decrease
the DEB, as will be shown in the simulations.}

\subsection{Problem Formulation}

\begin{figure}[t]
\centering{}\includegraphics[width=8cm]{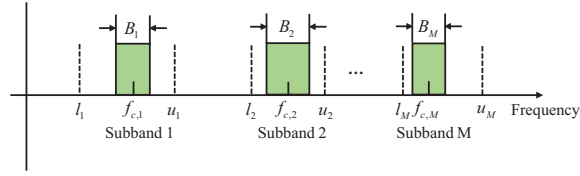}\caption{\label{fig:Op_dis}An illustration of the frequency distribution of
the multiband sensing systems with constraints.}
\end{figure}

At the transmitter of the multiband sensing systems, we aim to optimize
the system parameters for minimizing the delay SRL subject to a few
practical constraints. The optimization problem can be formulated
as\textcolor{blue}{
\begin{eqnarray}
\mathcal{P}: & \underset{\boldsymbol{\xi},\Delta\tau}{\min} & \Delta\tau(\boldsymbol{\xi})\nonumber \\
 & \text{s.t. } & \ensuremath{\sqrt{C_{\Delta\tau}(\Delta\tau,\boldsymbol{\xi})}=\Delta\tau},\label{eq:P2_SRL}\\
 &  & l_{m}\leq f_{c,m}-\frac{B_{m}}{2},\forall m,\label{eq:P2_box}\\
 &  & f_{c,m}\!+\!\frac{B_{m}}{2}\!\leq\!u_{m},\forall m,\label{eq:P2_box2}\\
 &  & f_{c,m}\!+\!\frac{B_{m}}{2}\!\leq\!f_{c,m+1}\!-\!\frac{B_{m+1}}{2},\!m\!=\!1,\!...,M\!-\!1,\label{eq:P2_notoverlap}\\
 &  & \sum_{m=1}^{M}B_{m}\leq W,\label{eq:P2_bandwidth}
\end{eqnarray}
}where $\boldsymbol{\xi}=[f_{c,1},\ldots,f_{c,M},N_{1},\ldots,N_{M}]^{T}$
denotes the vector consisting of system parameters needed to be optimized.\textcolor{blue}{{}
Choosing the carrier frequency $f_{c,m}$ and subcarrier number $N_{m}$
of each subband as the optimization variables is reasonable, since
in current communication standards (e.g., IEEE 802.11bf standard \cite{802.11bf_huawei}),
the subcarrier spacing $f_{s,m}$ is always fixed and only $f_{c,m}$
and $N_{m}$ are able to be optimized.} Note that $B_{m}=N_{m}f_{s,m}$
denotes the bandwidth of the $m$-th subband, the constraint (\ref{eq:P2_SRL})
is the definition of SRL, and (\ref{eq:P2_bandwidth}) is the total
bandwidth constraint, where $W$ denotes the maximum available bandwidth
for sensing over all subbands. The constraints (\ref{eq:P2_box})
and (\ref{eq:P2_box2}) are formulated to limit the frequency of each
subband in a given interval since only a few non-contiguous subbands
are available for sensing with limited bandwidths in practical communication
standards, where $l_{m}$ and $u_{m}$ denote the lower bound and
upper bound of the frequency for the $m$-th subband, respectively.
\textcolor{blue}{The constraints (\ref{eq:P2_notoverlap}) are formulated
to ensure that different subbands are not overlapped after optimization.}
The above mentioned parameters have been shown in Fig. \ref{fig:Op_dis}
for clarity.

\textcolor{blue}{Note that the definition of delay SRL only involves
two paths. However, the proposed optimization scheme can still be
applied to a practical multipath environment with more than two paths
and the effectiveness has been validated by simulation results. Specifically,
we can solve problem $\mathcal{P}$ with the SRL objective calculated
using the estimated mean amplitude $\tilde{\alpha}_{1}=\tilde{\alpha}_{2}=\frac{1}{\hat{K}}\sum_{k=1}^{\hat{K}}\left|\hat{\alpha}_{k}\right|$,
to get a solution $\boldsymbol{\xi}$ for the system parameter, where
the number of paths $\hat{K}$ and the amplitude for each path $\hat{\alpha}_{k}$
can be estimated in the previous time slot using a multiband delay
estimation algorithm, e.g., \cite{arxiv_SPVBI}.}

\subsection{Proposed Optimization Algorithm}

\textcolor{blue}{$\mathcal{P}$ is a mixed-integer nonlinear programming
(MINLP) problem, which involves the integer optimization variables
$N_{m},\forall m$ with the form of summation terms in the expression
of $C_{\Delta\tau}(\Delta\tau,\boldsymbol{\xi})$ (As can be seen
in (\ref{eq:FIM4})). Finding an optimal solution of a MINLP problem
is intractable because it is generally non-deterministic Polynomial-time
hard (NP-hard). Furthermore, the non-convex equality constraint in
$\mathcal{P}$ makes it more difficult to solve.}

\textcolor{blue}{To handle these issues, some optimization techniques
have been proposed to find a sub-optimal solution to the original
problem, such as convex relaxation techniques \cite{ConvexRelaxation,MINLP1},
metaheuristic techniques \cite{MINLP3,MINLP5}, and so on. However,
most methods can only find a local or approximate solution with high
computational complexity. To reduce complexity and make the problem
tractable, we consider a widely employed integer relaxation techniques
\cite{interior_point,Gradient}. Specifically, we first relax the
integer variable $N_{m}$ to a real variable and calculate $C_{\Delta\tau}(\Delta\tau,\boldsymbol{\xi})$
using the compact form of FIM in (\ref{eq:FIM_compact}), which transforms
the summation terms into a trigonometric product form. Then, we adopt
the AO algorithm to find a stationary point of $\mathcal{P}$ by alternatively
optimizing the variables of delay separation $\Delta\tau$ and system
parameters $\boldsymbol{\xi}$. Finally, we round down the results
of $N_{m}$ to obtain approximate integer solutions. The approximation
error caused by rounding down operation is acceptable, since it is
relatively negligible as compare to the absolute value of $N_{m}$,
which is always large in an OFDM system.}

\textcolor{blue}{For given $\boldsymbol{\xi}$, the subproblem is
to find a global minimum solution that satisfies the equality (\ref{eq:P2_SRL}):
\begin{eqnarray}
\mathcal{P}_{1}: & \underset{\Delta\tau}{\min} & \Delta\tau\nonumber \\
 & \text{s.t. } & \sqrt{C_{\Delta\tau}(\Delta\tau,\boldsymbol{\xi})}=\Delta\tau.\label{eq:P3}
\end{eqnarray}
}The solution can be easily found by one-dimensional search of $\Delta\tau$
with acceptable computational complexity. Then, for given $\Delta\tau$,
the subproblem transformed from $\mathcal{P}$ is given by
\begin{eqnarray}
\mathcal{P}_{2}: & \underset{\boldsymbol{\xi}}{\min} & C_{\Delta\tau}(\boldsymbol{\xi})\nonumber \\
 & \text{s.t. } & (\ref{eq:P2_box})-(\ref{eq:P2_bandwidth}).\label{eq:P2_bandwidth-1}
\end{eqnarray}
The subproblem $\mathcal{P}_{2}$ is a non-convex optimization problem
with linear inequality constraints and complicated non-convex objective
function $C_{\Delta\tau}(\boldsymbol{\xi})$. Thus we adopt the SCA
algorithm \cite{TOSCA,SCA} to find its stationary solution, which
iteratively updates $\boldsymbol{\xi}$ by solving a convex surrogate
problem obtained by replacing $C_{\Delta\tau}(\boldsymbol{\xi})$
with a convex surrogate function. Specifically, the SCA algorithm
contains three steps at each iteration as elaborated below.

\textbf{Step 1}: At the $t$-th iteration, by applying the first-order
Taylor expansion for $C_{\Delta\tau}(\boldsymbol{\xi})$, the surrogate
function is given by
\begin{equation}
\bar{f}^{t}(\boldsymbol{\xi})=f\left(\boldsymbol{\xi}^{t}\right)+\left(\mathbf{g}_{\boldsymbol{\xi}}^{t}\right)^{T}\left(\boldsymbol{\xi}-\boldsymbol{\xi}^{t}\right)+\omega\left\Vert \boldsymbol{\xi}-\boldsymbol{\xi}^{t}\right\Vert ^{2},\label{eq:SCA_surro_func}
\end{equation}
where $f\left(\boldsymbol{\xi}^{t}\right)=C_{\Delta\tau}(\boldsymbol{\xi}^{t})$,
$\omega>0$ is a constant, and $\mathbf{g}_{\boldsymbol{\xi}}^{t}=\partial_{\boldsymbol{\xi}}C_{\Delta\tau}(\boldsymbol{\xi}^{t})$
denotes the gradients, of which the $i$-th element is given by
\begin{equation}
\begin{aligned}g_{\xi_{i}}^{t} & =\frac{\partial C_{\Delta\tau}(\boldsymbol{\xi}^{t})}{\partial\xi_{i}}\\
 & =\frac{\partial(\mathbf{C}_{\boldsymbol{\eta}}^{t}(1,1)+\mathbf{C}_{\boldsymbol{\eta}}^{t}(2,2)-\mathbf{C}_{\boldsymbol{\eta}}^{t}(1,2)-\mathbf{C}_{\boldsymbol{\eta}}^{t}(2,1))}{\partial\xi_{i}}\\
 & =\mathbf{F}_{i}(1,1)+\mathbf{F}_{i}(2,2)-\mathbf{F}_{i}(1,2)-\mathbf{F}_{i}(2,1),
\end{aligned}
\label{eq:SCA_gradient}
\end{equation}
where $\mathbf{F}_{i}\triangleq-\mathbf{J}_{\boldsymbol{\eta}}^{-1}\frac{\partial\mathbf{J}_{\boldsymbol{\eta}}}{\partial\xi_{i}}\mathbf{J}_{\boldsymbol{\eta}}^{-1}$.

\textbf{Step 2}: In this step, the optimal solution $\bar{\boldsymbol{\xi}}^{t}$
of the following problem is obtained:
\begin{eqnarray}
\mathcal{P}_{2}^{'}: & \underset{\boldsymbol{\xi}}{\min} & \bar{f}^{t}(\boldsymbol{\xi})\nonumber \\
 & \text{s.t. } & (\ref{eq:P2_box})-(\ref{eq:P2_bandwidth}),\label{eq:P2_bandwidth-1-1}
\end{eqnarray}
which is a convex approximation of $\mathcal{P}_{2}$. Then, the convex
optimization problem $\mathcal{P}_{2}^{'}$ can be efficiently solved
by off-the-shelf solvers, e.g. the classical CVX solver.

\textbf{Step 3}: After obtaining $\bar{\boldsymbol{\xi}}^{t}$, $\boldsymbol{\xi}^{t}$
is updated according to
\begin{equation}
\boldsymbol{\xi}^{t+1}=\left(1-\sigma^{t}\right)\boldsymbol{\xi}^{t}+\sigma^{t}\bar{\boldsymbol{\xi}}^{t},\label{eq:SCA_update}
\end{equation}
where $\sigma^{t}$ is the step size determined by the Armijo rule
\cite{Bertsekas_book95_NProgramming}.

The proposed optimization algorithms are presented in Algorithm \ref{alg:AO}.
Note that to improve the probability of finding the global optimum,
we may repeatedly perform Algorithm \ref{alg:AO} with random initializations
and finally find the best solution. \textcolor{blue}{Finally, we prove
the convergence of the Algorithm \ref{alg:AO}.}
\begin{lem}
\textup{\textcolor{blue}{\label{lem:1}}}\textcolor{blue}{Let $\left\{ \boldsymbol{\xi}^{i},\Delta\tau^{i}\right\} _{i=1}^{\infty}$
denote the sequence of iterates generated by Algorithm 1. Then, $\left\{ \Delta\tau^{i}\right\} _{i=1}^{\infty}$
is a non-increasing sequence.}
\end{lem}
\textit{\textcolor{blue}{Proof}}\textcolor{blue}{: See Appendix \ref{subsec:convergence}.}

\textcolor{blue}{From Lemma \ref{lem:1}, the sequence $\left\{ \Delta\tau^{i}\right\} _{i=1}^{\infty}$
must converge to a limit $\Delta\tau^{*}>0$, i.e.,
\begin{equation}
\lim_{i\rightarrow\infty}\Delta\tau^{i}=\Delta\tau^{*}.\label{eq:limitingpoint}
\end{equation}
In the following theorem, we further prove that the sequence $\left\{ \boldsymbol{\xi}^{i},\Delta\tau^{i}\right\} _{i=1}^{\infty}$
converges to a stationary point that satisfies KKT conditions of the
following equivalent problem of $\mathcal{P}$:
\begin{eqnarray*}
\mathcal{P}_{e}: & \underset{\boldsymbol{\xi},\Delta\tau}{\min} & C_{\Delta\tau}(\Delta\tau,\boldsymbol{\xi})\\
 & \text{s.t. } & h\left(\boldsymbol{\xi},\Delta\tau\right)=0,\\
 &  & g_{j}\left(\boldsymbol{\xi}\right)\leq0,j=1,...,J,
\end{eqnarray*}
where $h\left(\boldsymbol{\xi},\Delta\tau\right)\triangleq\sqrt{C_{\Delta\tau}(\Delta\tau,\boldsymbol{\xi})}-\Delta\tau$
and $g_{j}\left(\boldsymbol{\xi}\right),\forall j$ represents all
inequality constraints in original problem $\mathcal{P}$.}
\begin{thm}
\textcolor{blue}{\label{thm:(Convergence-of-Algorithm}(Convergence
of Algorithm 1): Starting from a feasible initial point $\left\{ \boldsymbol{\xi}^{0},\Delta\tau^{0}\right\} $,
then sequence $\left\{ \boldsymbol{\xi}^{i},\Delta\tau^{i}\right\} _{i=1}^{\infty}$
generated by Algorithm \ref{alg:AO} has a limiting point which is
a KKT point of $\mathcal{P}_{e}.$}
\end{thm}
\textit{\textcolor{blue}{Proof}}\textcolor{blue}{: See Appendix \ref{subsec:convergence2}.}

\begin{algorithm}[t]
{\small{}\caption{\label{alg:AO}The multiband sensing system parameters optimization
algorithm}
}{\small\par}

\textbf{Input:} $f_{s,m}$,\textcolor{black}{{} $l_{m}$, $u_{m},\forall m$,
}$\boldsymbol{\alpha}$,\textcolor{black}{{} $\sigma_{ns}^{2}$, $W$,
maximum iteration number $I_{AO}$, $I_{SCA}$, threshold $\epsilon$.}

\textbf{Output:} $f_{c,m}^{*}$, $N_{m}^{*},\forall m$.

\begin{algorithmic}[1]

\STATE Initialize $f_{c,m}$, $N_{m},\forall m$.

\FOR{ $i=1,\cdots,I_{AO}$}

\STATE \textcolor{blue}{Given $\boldsymbol{\xi}^{i-1}$, solve the
problem $\mathcal{P}_{1}$.}

\FOR{${\color{blue}{\color{black}t=1,\cdots,I_{SCA}}}$}

\STATE \textcolor{blue}{Given $\Delta\tau^{i}$}, construct surrogate
functions based on (\ref{eq:SCA_surro_func}) and (\ref{eq:SCA_gradient}).

\STATE Find the optimal solution $\bar{\boldsymbol{\xi}}^{t}$ for
the problem $\mathcal{P}_{2}^{'}$.

\STATE Update the variables $\boldsymbol{\xi}^{t+1}$ based on (\ref{eq:SCA_update}).

\IF{\textcolor{black}{{} }$\left\Vert \boldsymbol{\xi}^{t+1}-\boldsymbol{\xi}^{t}\right\Vert \leq\epsilon$}

\STATE \textbf{\textcolor{black}{break}}

\ENDIF

\ENDFOR

\STATE\textcolor{blue}{{} $\boldsymbol{\xi}^{i}=\boldsymbol{\xi}^{t}$}

\ENDFOR

\end{algorithmic}
\end{algorithm}

\subsection{Simulation Results}

In this subsection, we provide numerical results to validate the effectiveness
of our proposed algorithms and give useful insights. The default system
parameters configurations are set as follows unless otherwise specified:
We consider that the measurements are collected at $M=2$ subbands
with subcarrier spacing $f_{s,1}=f_{s,2}=78.125$ KHz\textcolor{black}{.
The SNR} is set as $10$ dB and $K=2$ with complex scalars set as
$\alpha_{1}=0.8+0.6j$ and $\alpha_{2}=0.6+0.8j$. The overall bandwidth
constraint $W$ is $40$ MHz, and $l_{1},u_{1},l_{2},u_{2}=2.4,2.5,2.7,2.9$
GHz with an increasing order, respectively. \textcolor{black}{Note
that} we \textcolor{black}{deliberately} restrict our attention to
this simple case with equal subcarrier spacing at two subbands to
gain insights, although our formulated problem and proposed optimization
algorithms are applicable for \textcolor{black}{more practical scenarios}.

\begin{figure}[t]
\centering{}\textcolor{blue}{\includegraphics[width=7.2cm]{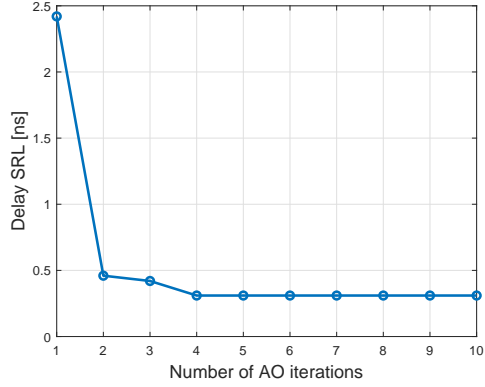}\caption{\textcolor{blue}{\label{fig:Op_converge}Convergence behavior of Algorithm
\ref{alg:AO}.}}
}
\end{figure}

We first illustrate the convergence behavior of the proposed Algorithm
\ref{alg:AO}. In Fig. \ref{fig:Op_converge}, we plot the delay SRL
versus the number of AO iterations. As can be seen, Algorithm \ref{alg:AO}
can converge within $5$ iterations rapidly.

\textcolor{blue}{}

\begin{figure}[t]
\centering{}\textcolor{blue}{\negthinspace{}\negthinspace{}\negthinspace{}\negthinspace{}}\subfloat[\label{fig:Op_SRL_SNR}]{\centering{}\textcolor{blue}{\includegraphics[width=4.9cm]{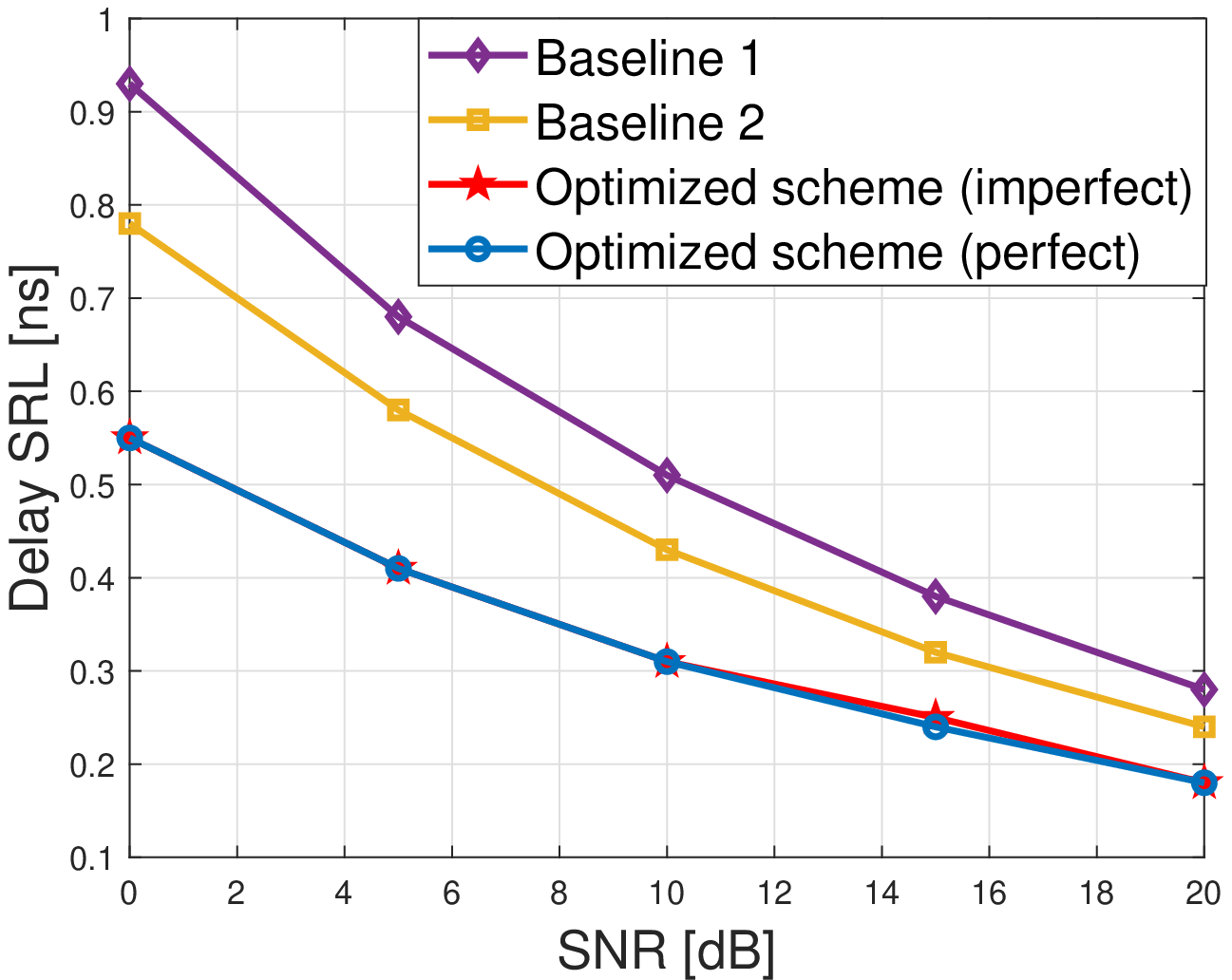}}}\textcolor{blue}{\negthinspace{}\negthinspace{}\negthinspace{}\negthinspace{}\negthinspace{}\negthinspace{}\negthinspace{}}\subfloat[\label{fig:Op_DEB_SNR}]{\centering{}\textcolor{blue}{\includegraphics[width=4.9cm]{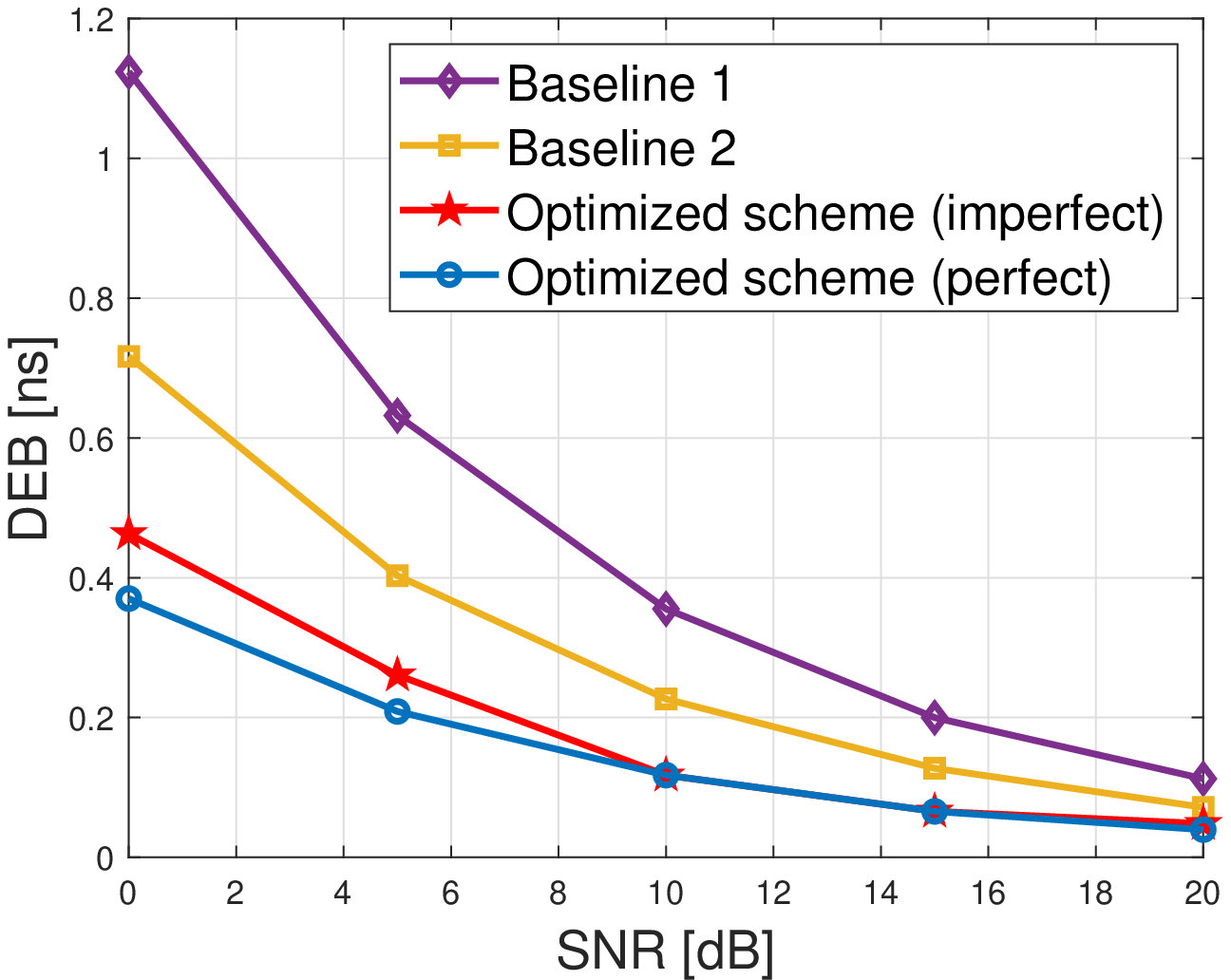}}}\textcolor{blue}{\caption{\textcolor{blue}{\label{fig:Op_SNR}An illustration of the SRL and
DEB versus $\textrm{SNR}$.}}
}
\end{figure}

In Fig. \ref{fig:Op_SRL_SNR}, we investigate the optimized results
of delay SRL versus the SNR. In particular, we consider two heuristic
baselines: (i) Baseline 1: Setting $f_{c,1}=\frac{l_{1}+u_{1}}{2}$,
$f_{c,2}=\frac{l_{2}+u_{2}}{2}$, and $B_{1}=B_{2}=W/2$; (ii) Baseline
2: Setting $f_{c,1}=l_{1}+B_{1}/2$, $f_{c,2}=u_{2}-B_{2}/2$, and
$B_{1}=B_{2}=W/2$.\textcolor{blue}{{} To verify the effectiveness of
the proposed optimization scheme using the estimated mean amplitude
information (imperfect), we compare the performance of the proposed
scheme with estimated and perfect amplitude information (perfect)
as well as the baselines in Fig. \ref{fig:Op_SRL_SNR}. It is observed
that: (1) The proposed schemes with imperfect and perfect information
achieve similar performance, indicating that the approximation errors
of $\alpha_{k}$'s have little impact on the final optimization results;
(2) The delay SRL decreases with the increase of SNR for all schemes;
(3) The optimized scheme reaps a large performance gain over the baselines
in the low and median SNR region; (4) Allocating the spectrum resource
uniformly among subbands (i.e., $B_{m}=W/M,\forall m$) is not the
optimal scheme.}

Fig. \ref{fig:Op_DEB_SNR} investigates DEB as a function of SNR based
on the optimized variables, which are obtained from the output of
Algorithm \ref{alg:AO}. As can be seen, the optimization scheme can
significantly decrease the DEB compared to the baselines, though the
objective of the optimization problem is to minimize the delay SRL
instead of DEB.

\textcolor{blue}{}

\begin{figure}[t]
\centering{}\includegraphics[width=7.2cm]{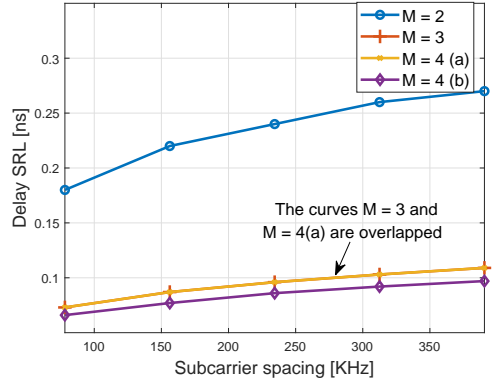}\caption{\label{fig:Op_fs}An illustration of SRL versus subcarrier spacing.}
\end{figure}

\begin{figure}[t]
\centering{}\includegraphics[width=7cm]{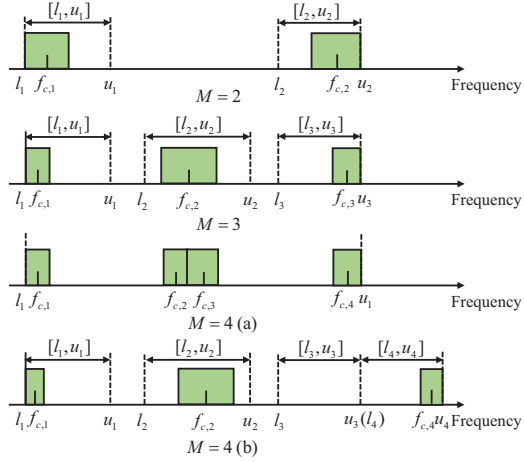}\caption{\label{fig:Op_fredistri}An illustration of the optimal frequency
band distribution.}
\end{figure}

Fig. \ref{fig:Op_fs} shows the SRL as a function of subcarrier spacing
for different number of subbands, where the bandwidth constraint $W=60$
MHz. The constrains (\ref{eq:P2_box}) and (\ref{eq:P2_box2}) are
set as in Table \ref{tab:Op_constraint}. It can be seen that the
SRL increases as the subcarrier spacing. Moreover, from the frequency
band distribution after optimization illustrated in Fig. \ref{fig:Op_fredistri}
and the optimal SRL shown in Fig \ref{fig:Op_fs}, we observe that:
\begin{enumerate}
\item The best delay SRL is obtained when the gap between the lowest frequency
point and the highest frequency point takes the maximum value, i.e.,
$f_{c,1}=l_{1}+\frac{B_{1}}{2}$ and $f_{c,M}=u_{M}-\frac{B_{M}}{2}$.
In fact, the gap is approximately equivalent to the largest frequency
band apertures, $f_{c,M}-f_{c,1}$. The delay SRL decreases with the
increase of the largest frequency band apertures, which is consistent
with the observations in Fig. \ref{fig:SRLvsFre}.
\item From the comparison of the cases $M=2$ and $M=3$, we observe that
interpolating a new subband between the existing two subbands can
significantly improve the resolution performance.
\item For the cases when $M\geq3$, the optimal spectrum allocation scheme
is similar, which divides the frequency band into three non-contiguous
subbands. Particularly, the middle subband is exactly at the central
of the subbands on both sides when it also satisfies the constraints
that lies in the feasible frequency intervals.
\end{enumerate}
\begin{table}[t]
\begin{centering}
\caption{\label{tab:Op_constraint}The setting of constraints in $\mathcal{P}_{2}$.}
\par\end{centering}
\centering{}%
\begin{tabular}{|>{\centering}m{1.3cm}|>{\centering}m{1.3cm}|>{\centering}m{1.3cm}|>{\centering}m{1.3cm}|>{\centering}m{1.3cm}|}
\hline 
 & {\footnotesize{}$[l_{1},u_{1}]$GHz} & {\footnotesize{}$[l_{2},u_{2}]$GHz} & {\footnotesize{}$[l_{3},u_{3}]$GHz} & {\footnotesize{}$[l_{4},u_{4}]$GHz}\tabularnewline
\hline 
\hline 
{\footnotesize{}$M=2$} & {\footnotesize{}$[2.4,2.5]$} & {\footnotesize{}$[3.1,3.2]$} & {\footnotesize{}\textbackslash} & {\footnotesize{}\textbackslash}\tabularnewline
\hline 
{\footnotesize{}$M=3$} & {\footnotesize{}$[2.4,2.5]$} & {\footnotesize{}$[2.7,2.9]$} & {\footnotesize{}$[3.1,3.2]$} & {\footnotesize{}\textbackslash}\tabularnewline
\hline 
{\footnotesize{}$M=4$ (a)} & {\footnotesize{}$[2.4,3.2]$} & {\footnotesize{}$[2.4,3.2]$} & {\footnotesize{}$[2.4,3.2]$} & {\footnotesize{}$[2.4,3.2]$}\tabularnewline
\hline 
{\footnotesize{}$M=4$ (b)} & {\footnotesize{}$[2.4,2.5]$} & {\footnotesize{}$[2.7,2.9]$} & {\footnotesize{}$[3.1,3.2]$} & {\footnotesize{}$[3.2,3.3]$}\tabularnewline
\hline 
\end{tabular}
\end{table}

\section{\label{sec:Conclusion}Conclusion}

In this paper, we studied the fundamental limits and optimization
of the multiband sensing systems. We derived a closed-form expression
of CRB for the delay separation and a corresponding theoretical analysis
is provided. We derived the SRL for the delay resolution and studied
the effect of frequency band apertures and phase distortions on the
SRL based on numerical results. We also derived performance bounds
CRB and ZZB for the delay estimation errors and give a comprehensive
performance analysis. Finally, we formulated a system parameters optimization
problem in the multiband sensing systems. An efficient algorithm has
been proposed to solve the non-convex optimization problem and useful
insights are provided based on numerical simulations.

\appendix

\subsection{\textcolor{blue}{\label{subsec:parameter identifiability} Proof
of Parameter Identifiability in (\ref{eq:refined_signal})}}

\textcolor{blue}{We consider two cases for the proof: (i) $\delta_{m}=0,\forall m$;
(ii) $\delta_{m}\neq0,\forall m$.}

\textcolor{blue}{(i) $\delta_{m}=0,\forall m$. In this case, the
signal model (\ref{eq:refined_signal}) can be rewritten as
\begin{equation}
y_{m,n}=\sum_{k=1}^{K}\!\alpha_{k}^{\prime}e^{-j2\pi f_{\!c,\!m}^{\prime}\!\tau_{k}}e^{-j2\pi nf_{\!s,\!m}\!\tau_{k}}e^{j\varphi_{m}^{\prime}}s_{m,n}+w_{m,n}.\label{eq:delta_0}
\end{equation}
}

\textcolor{blue}{Proving the parameter identifiability is equivalent
to prove that the \textquotedblleft identifiability equations\textquotedblright{}
of (\ref{eq:delta_0}) have a unique solution, which is given by
\begin{eqnarray}
\sum_{k=1}^{K}\!\check{\alpha}_{k}^{\prime}e^{-j2\pi f_{\!c,\!m}^{\prime}\!\check{\tau}_{k}}e^{-j2\pi nf_{\!s,\!m}\!\check{\tau}_{k}}e^{j\check{\varphi}_{m}^{\prime}} & = & \sum_{k=1}^{K}\!\alpha_{k}^{\prime}e^{-j2\pi f_{\!c,\!m}^{\prime}\!\tau_{k}}e^{-j2\pi nf_{\!s,\!m}\!\tau_{k}}e^{j\varphi_{m}^{\prime}},\label{eq:all_identifiability}\\
 &  & m=1,...,M,n=1,...,N_{m}.\nonumber 
\end{eqnarray}
We denote the unique solution of equality (\ref{eq:all_identifiability})
as: $\check{\alpha}_{k}^{\prime}=\alpha_{k}^{\prime}$, $\check{\tau}_{k}=\tau_{k}$,
$\check{\varphi}_{m}^{\prime}=\varphi_{m}^{\prime}$, $k=1,...,K$,
$m=2,...,M$. We first prove that $\check{\alpha}_{k}^{\prime}=\alpha_{k}^{\prime}$,
$\check{\tau}_{k}=\tau_{k}$ is the unique solution of (\ref{eq:all_identifiability}).
Then, we prove that $\check{\varphi}_{m}^{\prime}=\varphi_{m}^{\prime}$
is also the unique solution of (\ref{eq:all_identifiability}) to
complete the proof.}

\textcolor{blue}{When $m=1$, the \textquotedblleft identifiability
equations\textquotedblright{} are given by
\begin{eqnarray*}
\sum_{k=1}^{K}\check{\alpha}_{k}^{\prime}e^{-j2\pi nf_{s,1}\check{\tau}_{k}} & = & \sum_{k=1}^{K}\alpha_{k}^{\prime}e^{-j2\pi nf_{s,1}\tau_{k}},n=1,...,N_{1},
\end{eqnarray*}
which can be equivalently transformed to a linear form
\begin{equation}
\mathbf{\check{A}}\boldsymbol{\check{\beta}}=\mathbf{A}\boldsymbol{\beta},\label{eq:identifiability}
\end{equation}
where
\[
\mathbf{A}=\left[\begin{array}{ccc}
e^{-j2\pi(-\frac{N_{1}-1}{2})f_{s,1}\tau_{1}} & \ldots & e^{-j2\pi(-\frac{N_{1}-1}{2})f_{s,1}\tau_{K}}\\
\vdots & \ddots & \vdots\\
e^{-j2\pi(\frac{N_{1}-1}{2})f_{s,1}\tau_{1}} & \ldots & e^{-j2\pi(\frac{N_{1}-1}{2})f_{s,1}\tau_{K}}
\end{array}\right]\in\mathbb{C}^{N_{1}\times K},
\]
$\boldsymbol{\beta}=\left[\alpha_{1}^{\prime},\ldots,\alpha_{K}^{\prime}\right]^{T}$,
$\mathbf{\check{A}}$ and $\boldsymbol{\check{\beta}}$ have similar
definitions. Then, we have the following theorem:}
\begin{thm}
\textcolor{blue}{$\check{\alpha}_{k}^{\prime}=\alpha_{k}^{\prime}$,
$\check{\tau}_{k}=\tau_{k}$ is the unique solution of (\ref{eq:identifiability})
if $N_{1}+1>2K$.}
\end{thm}
\textit{\textcolor{blue}{Proof}}\textcolor{blue}{: See \cite{Nehorai1991,Li2007}.}

\textcolor{blue}{Apparently, the unique solution $\check{\alpha}_{k}^{\prime}=\alpha_{k}^{\prime}$,
$\check{\tau}_{k}=\tau_{k}$ in (\ref{eq:identifiability}) is also
the unique solution for all identifiability equations in (\ref{eq:all_identifiability}).
Then, we prove that (\ref{eq:all_identifiability}) has a unique solution:
$\check{\varphi}_{m}^{\prime}=\varphi_{m}^{\prime}$. The identifiability
equations of the $m$-th subband can be written as
\begin{equation}
\check{\boldsymbol{\varPhi}}\check{\boldsymbol{b}}=\boldsymbol{\varPhi}\boldsymbol{b},\label{eq:identifiability_M}
\end{equation}
where $\boldsymbol{b}\in\mathbb{C}^{N_{m}\times1}$ whose $n$-th
element is $\sum_{k=1}^{K}\!\alpha_{k}^{\prime}e^{-j2\pi f_{\!c,\!m}^{\prime}\!\tau_{k}}e^{-j2\pi nf_{\!s,\!m}\!\tau_{k}}$,
$\boldsymbol{\varPhi}=\textrm{diag}(e^{j\varphi_{m}^{\prime}},\ldots,e^{j\varphi_{m}^{\prime}})\in\mathbb{C}^{N_{m}\times N_{m}}$,
$\check{\boldsymbol{\varPhi}}$ and $\check{\boldsymbol{b}}$ have
similar definitions. Due to the uniqueness of $\check{\alpha}_{k}^{\prime}$
and $\check{\tau}_{k}$ proved as above, we have $\check{\boldsymbol{b}}=\boldsymbol{b}$.
Hence, we have a unique solution of (\ref{eq:identifiability_M}),
i.e., $\check{\varphi}_{m}^{\prime}=\varphi_{m}^{\prime}$. Similarly,
for identifiability equations of other subbands, the unique solution
can also be guaranteed, i.e., $\check{\varphi}_{m}^{\prime}=\varphi_{m}^{\prime},m=2,...,M$,
which completes the proof.}

\textcolor{blue}{(ii) $\delta_{m}\neq0,\forall m$. In this case,
the signal model is ambiguous and the parameters to be estimated are
non-identifiable. One approach to eliminate the ambiguity is to introduce
a prior distribution of the parameters. In this way, the estimation
problem becomes a Bayesian one. In our case, the parameter $\delta_{m}$
follows a prior distribution $p\left(\delta_{m}\right)\sim\mathcal{N}\left(0,\sigma_{p}^{2}\right)$,
where $\sigma_{p}$ is assumed to be small. Consequently, the prior
information of $\boldsymbol{\delta}$ is helpful to eliminate the
ambiguity of signal model (\ref{eq:refined_signal}) and thus the
estimation performance mainly depends on the prior information precision
$(\sigma_{p}^{2})^{-1}$.}

\textcolor{blue}{In summary, when $\delta_{m}=0,\forall m$, the parameters
to be estimated in (\ref{eq:refined_signal}) are identifiable under
a condition that $N_{1}+1>2K$. When $\delta_{m}\neq0,\forall m$,
though the parameters to be estimated in (\ref{eq:refined_signal})
are non-identifiable, the prior information of $\boldsymbol{\delta}$
can help to eliminate the ambiguity of signal model (\ref{eq:refined_signal}).
The sensing performance can still be guaranteed with a small $\sigma_{p}$.}

\subsection{\label{subsec:FIM_ori}Elements in (\ref{eq:FIM})}

We define $\overline{\mathbf{J}}_{w}=\mathbb{E}_{\boldsymbol{y}}\left[-\frac{\partial^{2}\ln f(\boldsymbol{y}\mid\boldsymbol{\eta})}{\partial\boldsymbol{\eta}\partial\boldsymbol{\eta}^{T}}\right]$,
whose elements are given by
\begin{equation}
\overline{\Psi}\left(x_{r},x_{s}\right)=\frac{2}{\sigma_{ns}^{2}}\sum_{m,n}\textrm{Re}\left\{ \frac{\partial\boldsymbol{\mu}^{H}}{\partial x_{r}}\frac{\partial\boldsymbol{\mu}}{\partial x_{s}}\right\} .\label{eq:FIM2}
\end{equation}
Then, the elements of $\mathbf{J}_{w}$ are obtained by taking the
expectation of $\overline{\mathbf{J}}_{w}$ over the random parameter
$\boldsymbol{\delta}$. Besides, the FIM $\mathbf{J}_{p}$ is given
by
\begin{equation}
\mathbf{J}_{p}=\textrm{diag}\left(0,\ldots,0,1/\sigma_{p}^{2},\ldots,1/\sigma_{p}^{2}\right),\label{eq:FIM3}
\end{equation}
where only the diagonal elements associated with the position of block
matrix $\Psi(\boldsymbol{\delta},\boldsymbol{\delta})$ are non-zero.
Finally, based on (\ref{eq:FIM1}), (\ref{eq:FIM2}), and (\ref{eq:FIM3}),
the entries of $\mathbf{J}_{\boldsymbol{\eta}}$ are given by
\[
\begin{aligned} & \Psi\left(\tau_{r},\tau_{s}\right)=\frac{8\pi^{2}}{\sigma_{ns}^{2}}\sum_{m,n}f_{m,n}^{2}\textrm{Re}\left\{ (\alpha_{r}^{\prime})^{*}\alpha_{s}^{\prime}e^{j2\pi f_{m,n}\left(\tau_{r}-\tau_{s}\right)}\right\} ,\\
 & \Psi\left(\tau_{r},\alpha_{R,s}\right)=\frac{4\pi}{\sigma_{ns}^{2}}\sum_{n,m}\textrm{Re}\left\{ jf_{m,n}(\alpha_{r}^{\prime})^{*}e^{j2\pi f_{m,n}\left(\tau_{r}-\tau_{s}\right)}\right\} ,\\
 & \Psi\left(\tau_{r},\alpha_{I,s}\right)=\frac{-4\pi}{\sigma_{ns}^{2}}\sum_{n,m}\textrm{Re}\left\{ f_{m,n}(\alpha_{r}^{\prime})^{*}e^{j2\pi f_{m,n}\left(\tau_{r}-\tau_{s}\right)}\right\} ,\\
 & \Psi\left(\tau_{r},\varphi_{i}^{\prime}\right)=\!-\frac{4\pi}{\sigma_{ns}^{2}}\!\sum_{n,m=i}\!\!\!\textrm{Re}\!\left\{ \!f_{i,n}(\alpha_{r}^{\prime})^{*}\!\!\sum_{k=1}^{K}\alpha_{k}e^{j2\pi f_{i,n}\left(\tau_{r}-\tau_{k}\right)}\!\right\} \!,\\
 & \Psi\left(\tau_{r},\delta_{i}\right)\!=\!\frac{8\pi^{2}}{\sigma_{ns}^{2}}\!\!\sum_{n,m=i}\!\!\!\!nf_{i,n}f_{s,i}\textrm{Re}\!\left\{ \!\!(\alpha_{r}^{\prime})^{*}\!\!\sum_{k=1}^{K}\!\alpha_{k}^{\prime}e^{j2\pi f_{i,n}\left(\!\tau_{r}-\tau_{k}\!\right)}\!\!\right\} \!,\\
 & \Psi\left(\alpha_{R,r},\alpha_{R,s}\right)=\Psi\left(\alpha_{I,r},\alpha_{I,s}\right)\\
 & \quad=\frac{2}{\sigma_{ns}^{2}}\sum_{m,n}\cos(2\pi f_{m,n}(\tau_{r}-\tau_{s})),\\
 & \Psi\left(\alpha_{R,r},\alpha_{I,s}\right)=-\Psi\left(\alpha_{I,r},\alpha_{R,s}\right)\\
 & \quad=-\frac{2}{\sigma_{ns}^{2}}\sum_{m,n}\sin(2\pi f_{m,n}(\tau_{r}-\tau_{s})),
\end{aligned}
\]

\begin{equation}
\begin{aligned} & \Psi\left(\alpha_{R,r},\varphi_{i}^{\prime}\right)=\frac{2}{\sigma_{ns}^{2}}\!\sum_{n,m=i}\!\textrm{Re}\left\{ j\sum_{k=1}^{K}\alpha_{k}^{\prime}e^{j2\pi f_{i,n}\left(\tau_{r}-\tau_{k}\right)}\right\} ,\\
 & \Psi\left(\alpha_{R,r},\delta_{i}\right)\!=\!-\frac{4\pi}{\sigma_{ns}^{2}}\!\sum_{n,m=i}\!\textrm{Re}\!\left\{ \!jnf_{s,i}\sum_{k=1}^{K}\alpha_{k}^{\prime}e^{j2\pi f_{n,i}\left(\tau_{r}-\tau_{k}\right)}\!\right\} \!,\\
 & \Psi\left(\alpha_{I,r},\varphi_{i}\right)=\frac{2}{\sigma_{ns}^{2}}\sum_{n,m=i}\textrm{Re}\left\{ \sum_{k=1}^{K}\alpha_{k}^{\prime}e^{j2\pi f_{i,n}\left(\tau_{r}-\tau_{k}\right)}\right\} ,\\
 & \Psi\left(\alpha_{I,r},\delta_{i}\right)\!=\!-\frac{4\pi}{\sigma_{ns}^{2}}\!\sum_{n,m=i}\!\textrm{Re}\left\{ \!nf_{s,i}\sum_{k=1}^{K}\alpha_{k}^{\prime}e^{j2\pi f_{i,n}\left(\tau_{r}-\tau_{k}\right)}\right\} \!,\\
 & \Psi\left(\varphi_{r}^{\prime},\varphi_{s}^{\prime}\right)=\left\{ \!\begin{array}{c}
\frac{2}{\sigma_{ns}^{2}}\!\underset{n,m=r}{\sum}\left|\sum_{k=1}^{K}\alpha_{k}^{\prime}e^{-j2\pi f_{r,n}\tau_{k}}\right|^{2}\!,r=s\\
0,\text{otherwise},
\end{array}\right.\\
 & \Psi\left(\varphi_{r}^{\prime},\delta_{s}\right)=\left\{ \!\!\!\begin{array}{c}
-\frac{4\pi}{\sigma_{ns}^{2}}\!\underset{n,m=r}{\sum}\!\!nf_{s,r}\!\left|\sum_{k=1}^{K}\alpha_{k}^{\prime}e^{-j2\pi f_{r,n}\tau_{k}}\right|^{2}\!\!,\!r=s\\
0,\text{otherwise},
\end{array}\right.\\
 & \Psi\left(\delta_{r},\delta_{s}\right)=\left\{ \!\!\!\begin{array}{c}
\frac{8\pi^{2}}{\sigma_{ns}^{2}}\!\underset{n,m=r}{\sum}\!\!\!\!n^{2}f_{s,r}^{2}\!\left|\sum_{k=1}^{K}\!\alpha_{k}^{\prime}e^{-j2\pi f_{r,n}\tau_{k}}\right|^{2}\!\!\!+\!\frac{1}{\sigma_{p}^{2}}\!,r\!=\!s\\
0,\text{otherwise}.
\end{array}\right.
\end{aligned}
\label{eq:FIM4}
\end{equation}

\subsection{\label{subsec:B.Compactly FIM}The Compactly Reformulated FIM of
(\ref{eq:FIM4})}

The Dirichlet kernel is given by
\begin{equation}
s(x)=\sum_{n=-(N-1)/2}^{(N-1)/2}e^{jnx}=\frac{\sin\left(\frac{N}{2}x\right)}{\sin\left(\frac{1}{2}x\right)}.\label{eq:dirichlet_kernel}
\end{equation}
We denote $\alpha_{k}^{\prime}=a_{k}e^{j\phi_{k}},\forall k$, where
$a_{k}$ and $\phi_{k}$ are the amplitude and phase of $\alpha_{k}^{\prime}$,
respectively. Then the reformulated FIM can be derived based on (\ref{eq:dirichlet_kernel}),
which is given by
\[
\begin{aligned} & \Psi\left(\tau_{r},\tau_{s}\right)=\frac{2}{\sigma_{ns}^{2}}\sum_{m=1}^{M}a_{r}a_{s}[4\pi^{2}f_{c,m}^{2}\cos(\psi_{m}-\Delta\phi)\gamma_{m}\\
 & \quad\quad\quad\quad\quad+\!4\pi f_{c,m}\sin(\psi\!-\!\Delta\phi)\gamma_{m}^{\prime}\!-\!\cos(\psi_{m}\!-\!\Delta\phi)\gamma_{m}^{\prime\prime}],\\
 & \Psi\left(\tau_{r},\alpha_{R,s}\right)=\frac{2}{\sigma_{ns}^{2}}\sum_{m=1}^{M}2\pi a_{r}f_{c,m}\sin(\psi_{m}+\phi_{r})\gamma_{m}\\
 & \quad\quad\quad\quad\quad\quad-a_{r}\cos(\psi_{m}+\phi_{r})\gamma_{m}^{\prime},\\
 & \Psi\left(\tau_{r},\alpha_{I,s}\right)=\frac{2}{\sigma_{ns}^{2}}\sum_{m=1}^{M}-2\pi a_{r}f_{c,m}\cos(\psi_{m}+\phi_{r})\gamma_{m}\\
 & \quad\quad\quad\quad\quad\quad-a_{r}\sin(\psi_{m}+\phi_{r})\gamma_{m}^{\prime},\\
 & \Psi\left(\alpha_{R,r},\alpha_{R,s}\right)=\Psi\left(\alpha_{I,r},\alpha_{I,s}\right)=\frac{2}{\sigma_{ns}^{2}}\sum_{m=1}^{M}\cos(\psi_{m})\gamma_{m},\\
 & \Psi\left(\alpha_{R,r},\alpha_{I,s}\right)=-\Psi\left(\alpha_{R,s},\alpha_{I,r}\right)=\frac{2}{\sigma_{ns}^{2}}\sum_{m=1}^{M}\sin(\psi_{m})\gamma_{m},\\
 & \Psi\left(\tau_{r},\varphi_{2}^{\prime}\right)=-\frac{2}{\sigma_{ns}^{2}}[2\pi f_{c,2}(a_{r}^{2}N_{2}+a_{1}a_{2}\cos(\psi_{2}-\Delta\phi)\gamma_{2})\\
 & \quad\quad\quad\quad\quad+a_{1}a_{2}\sin(\psi_{2}-\Delta\phi)\gamma_{2}^{\prime}],\\
 & \Psi\left(\alpha_{R,r},\varphi_{m}^{\prime}\right)=\frac{2}{\sigma_{ns}^{2}}(-N_{m}a_{r}\sin(\phi_{r})\\
 & \quad\quad\quad\quad\quad\quad\quad-a_{s}\sin(\psi_{m}+\phi_{s})\gamma_{m}),r=1,2;s\neq r,\\
 & \Psi\left(\alpha_{I,r},\varphi_{m}^{\prime}\right)=\frac{2}{\sigma_{ns}^{2}}(N_{m}a_{r}\cos(\phi_{r})+a_{s}\cos(\psi_{m}+\phi_{s})\gamma_{m},\\
 & \Psi\left(\varphi_{2}^{\prime},\varphi_{2}^{\prime}\right)=\frac{2}{\sigma_{ns}^{2}}(N_{2}(a_{1}^{2}+a_{2}^{2})+a_{1}a_{2}\gamma_{2}(\cos(\psi_{2}+\phi_{1})\\
 & \quad\quad\quad\quad\quad\quad+\cos(\psi_{2}-\phi_{2}))),\\
 & \Psi\left(\varphi_{2}^{\prime},\delta_{2}\right)=\frac{2}{\sigma_{ns}^{2}}(-a_{1}a_{2}\gamma_{m}^{\prime}(\sin(\psi_{m}\!+\!\phi_{1})+\sin(\psi_{m}\!-\!\phi_{2}))),
\end{aligned}
\]

\begin{equation}
\begin{aligned} & \Psi\left(\tau_{r},\delta_{m}\right)=\frac{2}{\sigma_{ns}^{2}}(4\pi^{2}a_{r}^{2}f_{s,m}^{2}\frac{N_{m}^{3}-N_{m}}{12}-a_{1}a_{2}\\
 & \quad\quad\quad\quad\quad\times\!\left[6\pi f_{c,m}\sin(\psi_{m}\!-\!\Delta\phi)\gamma_{m}^{\prime}\!\!+\!\cos(\psi_{m}\!-\!\Delta\phi)\gamma_{m}^{\prime\prime}\right]),\\
 & \Psi\left(\alpha_{R,r},\delta_{m}\right)=\frac{2}{\sigma_{ns}^{2}}(-a_{s}\cos(\psi_{m}+\phi_{s})\gamma_{m}^{\prime}),\\
 & \Psi\left(\alpha_{I,r},\delta_{m}\right)=\frac{2}{\sigma_{ns}^{2}}(-a_{s}\sin(\psi_{m}+\phi_{s})\gamma_{m}^{\prime}),\\
 & \Psi\left(\delta_{m},\delta_{m}\right)=\frac{2}{\sigma_{ns}^{2}}[4\pi^{2}(a_{1}^{2}+a_{2}^{2})\frac{N_{m}^{3}-N_{m}}{12}f_{s,m}^{2}-a_{1}a_{2}\gamma_{m}^{\prime\prime}\\
 & \quad\quad\quad\quad\quad\times(\cos(\psi_{m}+\phi_{1})+\cos(\psi_{m}-\phi_{2}))]+\frac{1}{\sigma_{p}^{2}},
\end{aligned}
\label{eq:FIM_compact}
\end{equation}
where $\gamma_{m}=\frac{\sin(\pi N_{m}f_{s,m}\Delta\tau)}{\sin(\pi f_{s,m}\Delta\tau)}$,
$\gamma_{m}^{\prime}=\frac{\partial\gamma_{m}}{\partial\Delta\tau}$,
$\gamma_{m}^{\prime\prime}=\frac{\partial^{2}\gamma_{m}}{\partial(\Delta\tau)^{2}}$,
$\psi_{m}=2\pi f_{c,m}\Delta\tau$, and $\Delta\phi=\phi_{2}-\phi_{1}$.

\subsection{\textcolor{blue}{\label{subsec:CRB_Closed_APP}The derivation for
$\textrm{CRB}_{\textrm{up}}$ and $\textrm{CRB}_{\textrm{low}}$ in
(\ref{eq:CRBup_low})}}

\textcolor{blue}{First, we present the closed-form expression of $C_{\Delta\tau}$,
which is given by
\begin{equation}
C_{\Delta\tau}(t)\triangleq f(t)=\frac{a+bt}{ct^{2}+dt+e},\label{eq:C_deltatau}
\end{equation}
where $t=\cos(2\pi\Delta f_{c}\Delta\tau)\in[-1,1]$ is a carrier
term, $a,b,c,d,e$ are intricate coefficients given by
\[
\begin{aligned}a= & 12\overline{N}^{2}-6\mathit{\gamma}^{2},\\
b= & -6\mathit{\gamma}^{2},\\
c= & -3\mathit{\gamma}^{2}\gamma^{\prime\prime}+3\gamma\mathit{\gamma^{\prime}}^{2},\\
d= & 6\pi^{2}\Delta f_{c}^{2}\gamma^{3}+(-2\overline{N}^{3}\pi^{2}f_{s}^{2}+2\overline{N}\pi^{2}f_{s}^{2}-6\gamma^{\prime\prime})\mathit{\gamma}^{2}\\
 & +(-6\overline{N}^{2}\pi^{2}\Delta f_{c}^{2}+6\mathit{\gamma^{\prime}}^{2})\gamma+6\overline{N}^{2}\gamma^{\prime\prime}-6\overline{N}\mathit{\gamma^{\prime}}^{2},\\
e= & 4\overline{N}^{5}\pi^{2}f_{s}^{2}+\pi^{2}(-2\mathit{f_{s}}^{2}\mathit{\gamma}^{2}+12\mathit{\Delta f_{c}}^{2}-4\mathit{f_{s}^{2}})\overline{N}^{3}+(-6\pi^{2}\mathit{\Delta f_{c}}^{2}\mathit{\gamma}+6\mathit{\gamma^{\prime\prime}})\overline{N}^{2}\\
 & +((-12\mathit{\Delta f_{c}}^{2}+2f_{s}^{2})\mathit{\gamma}^{2}\pi^{2}-6\mathit{\gamma^{\prime}}^{2})\overline{N}-3\mathit{\gamma}^{2}\gamma^{\prime\prime}+3\gamma\mathit{\gamma^{\prime}}^{2},
\end{aligned}
\]
When $\overline{N}\rightarrow\infty$, from the expressions of $\gamma,\gamma^{\prime}$,
and $\gamma^{\prime\prime}$, it can be shown that $\gamma,\gamma^{\prime}$,
and $\gamma^{\prime\prime}$ are bounded. Hence, when $\overline{N}f_{s}\Delta\tau$
is not an integer, we have $\gamma\neq0$ and}

\textcolor{blue}{
\begin{equation}
f(t)=\frac{a}{\overline{d}t+e}+\varepsilon\left(\overline{N}\right),
\end{equation}
where $\overline{d}=-2\overline{N}^{3}\pi^{2}f_{s}^{2}\mathit{\gamma}^{2}<0$
and $\frac{\varepsilon\left(\overline{N}\right)\left(\overline{d}t+e\right)}{a}\rightarrow0$
as $\overline{N}\rightarrow\infty$$.$ Therefore, $f(t)$ is a monotonic
function for $t\in[-1,1]$ when $\overline{N}$ is sufficiently large
and $\overline{N}f_{s}\Delta\tau$ is not an integer. Finally, the
upper (lower) bound of $C_{\Delta\tau}$, i.e., $\textrm{CRB}_{\textrm{up}}$
and $\textrm{CRB}_{\textrm{low}}$ in (\ref{eq:CRBup_low}), can be
obtained by taking the values $t=\pm1$.}

\subsection{\textcolor{blue}{\label{subsec:convergence}Proof of Lemma \ref{lem:1}}}

\textcolor{blue}{For the $i$-th iteration, we have $C_{\Delta\tau}(\boldsymbol{\xi}^{i},\Delta\tau^{i})\leq C_{\Delta\tau}(\boldsymbol{\xi}^{i-1},\Delta\tau^{i})$.
This inequality holds because the adopted SCA algorithm starts from
the initial point $\boldsymbol{\xi}^{i-1},\Delta\tau^{i}$ and decreases
$C_{\Delta\tau^{i}}(\boldsymbol{\xi})$ until converging to a stationary
point of $\mathcal{P}_{2}$. Then, we have
\begin{equation}
\sqrt{C_{\Delta\tau}(\boldsymbol{\xi}^{i},\Delta\tau^{i})}\leq\sqrt{C_{\Delta\tau}(\boldsymbol{\xi}^{i-1},\Delta\tau^{i})}\overset{a}{=}\Delta\tau^{i},\label{eq:Lemma1_1}
\end{equation}
where (\ref{eq:Lemma1_1}-$a$) holds due to that when we solve the
problem $\mathcal{P}_{1}$ with $\boldsymbol{\xi}=\boldsymbol{\xi}^{i-1}$
to get the solution $\Delta\tau^{i}$, $(\boldsymbol{\xi}^{i-1},\Delta\tau^{i})$
must satisfy the equality constraint in $\mathcal{P}_{1}$. Then,
from the expression of CRB in (\ref{eq:C_delattau}) and the corresponding
FIM expression (\ref{eq:FIM_compact}), it can be shown that there
always exists a sufficiently small number $\varepsilon_{\Delta\tau}>0$
such that
\begin{equation}
\sqrt{C_{\Delta\tau}(\boldsymbol{\xi}^{i},\varepsilon_{\Delta\tau})}>\varepsilon_{\Delta\tau}.\label{eq:Lemma1_2}
\end{equation}
Next, based on (\ref{eq:Lemma1_1}) and (\ref{eq:Lemma1_2}), we have
\begin{equation}
(\sqrt{C_{\Delta\tau}(\boldsymbol{\xi}^{i},\Delta\tau^{i})}-\Delta\tau^{i})(\sqrt{C_{\Delta\tau}(\boldsymbol{\xi}^{i},\varepsilon_{\Delta\tau})}-\varepsilon_{\Delta\tau})\leq0.
\end{equation}
Then, since $C_{\Delta\tau}$ is a continuous function and according
to the Existence Theorem of Zero Points, we can obtain a zero point
$\zeta\in(\varepsilon_{\Delta\tau},\Delta\tau^{i}]$ satisfying the
equality
\begin{equation}
\sqrt{C_{\Delta\tau}(\boldsymbol{\xi}^{i},\zeta)}-\zeta=0.\label{eq:zero_point}
\end{equation}
We know that $\Delta\tau^{i+1}$ is the global optimum of $\mathcal{P}_{1}$
at the ($i+1$)-th iteration, which means that it is also the minimum
zero point of (\ref{eq:zero_point}). Hence, we finally have $\Delta\tau^{i+1}\leq\zeta\leq\Delta\tau^{i}$.}

\subsection{\textcolor{blue}{\label{subsec:convergence2}Proof of Theorem \ref{thm:(Convergence-of-Algorithm}}}

\textcolor{blue}{We called $\boldsymbol{\xi}^{*},\Delta\tau^{*}$
a stationary point of $\mathcal{P}_{e}$/$\mathcal{P}$ if it satisfies
the KKT condition of $\mathcal{P}_{e}$, i.e., there exists $\mu_{1},...,\mu_{J}$
and $\lambda$ such that}

\textcolor{blue}{
\begin{eqnarray}
\nabla C_{\Delta\tau}(\boldsymbol{\xi}^{*},\Delta\tau^{*})+\sum_{j=1}^{J}\mu_{j}\nabla g_{j}\left(\boldsymbol{\xi}^{*}\right)+\lambda\nabla h\left(\boldsymbol{\xi}^{*},\Delta\tau^{*}\right) & = & 0,\label{eq:OriKKT_1}\\
\mu_{j}g_{j}\left(\boldsymbol{\xi}^{*}\right) & = & 0,\forall j,\label{eq:OriKKT_2}\\
g_{j}\left(\boldsymbol{\xi}^{*}\right) & \leq & 0,\forall j,\label{eq:OriKKT_3}\\
\mu_{j} & \geq & 0,\forall j,\label{eq:OriKKT_4}\\
h\left(\boldsymbol{\xi}^{*},\Delta\tau^{*}\right) & = & 0,\label{eq:OriKKT_5}
\end{eqnarray}
where $\nabla$ denotes the gradient operator. Subsequently, we would
like to prove that the limiting point $\boldsymbol{\xi}^{*},\Delta\tau^{*}$
of the sequence $\left\{ \boldsymbol{\xi}^{i},\Delta\tau^{i}\right\} _{i=1}^{\infty}$
satisfies the KKT conditions (\ref{eq:OriKKT_1})-(\ref{eq:OriKKT_5}).}

\textcolor{blue}{For the $i$-th iteration, the stationary point of
problem $\mathcal{P}_{2}$ obtained from the SCA algorithm satisfies
the KKT condition of $\mathcal{P}_{2}$, i.e.,
\begin{equation}
\begin{aligned} & \nabla C_{\Delta\tau}(\boldsymbol{\xi}^{i},\Delta\tau^{i})+\sum_{j=1}^{J}\tilde{\mu}_{j}\nabla g_{j}\left(\boldsymbol{\xi}^{i}\right)=0,\\
 & g_{j}\left(\boldsymbol{\xi}^{i}\right)\leq0,\forall j,\\
 & \tilde{\mu}_{j}\geq0,\forall j,\\
 & \tilde{\mu}_{j}g_{j}\left(\boldsymbol{\xi}^{i}\right)=0,\forall j.
\end{aligned}
\label{eq:P4SCA_KKT}
\end{equation}
Let $\lambda=0,\mu_{j}=\tilde{\mu}_{j},\forall j,$ according to (\ref{eq:P4SCA_KKT}),
we have the limiting point $\boldsymbol{\xi}^{*},\Delta\tau^{*}$
satisfying the KKT conditions (\ref{eq:OriKKT_1})-(\ref{eq:OriKKT_4})
of problem $\mathcal{P}_{e}$. Then, we prove that the limiting point
also satisfies the KKT condition (\ref{eq:OriKKT_5}). We have
\begin{equation}
\begin{aligned}\lim_{i\rightarrow\infty}\sqrt{C_{\Delta\tau}(\boldsymbol{\xi}^{i},\Delta\tau^{i})} & \overset{a}{=}\lim_{i\rightarrow\infty}\sqrt{C_{\Delta\tau}(\boldsymbol{\xi}^{i},\Delta\tau^{i+1})}\\
 & =\lim_{i\rightarrow\infty}\Delta\tau^{i+1}\\
 & \overset{b}{=}\Delta\tau^{i},
\end{aligned}
\label{eq:h=00003D0}
\end{equation}
where (\ref{eq:h=00003D0}-$a$) and (\ref{eq:h=00003D0}-$b$) follow
from the continuity of function $\sqrt{C_{\Delta\tau}(\boldsymbol{\xi},\Delta\tau)}$
and (\ref{eq:limitingpoint}). Therefore, following from (\ref{eq:h=00003D0}),
the limiting point $\boldsymbol{\xi}^{*},\Delta\tau^{*}$ of the sequence
$\left\{ \boldsymbol{\xi}^{i},\Delta\tau^{i}\right\} _{i=1}^{\infty}$
satisfies the KKT condition (\ref{eq:OriKKT_5}), which completes
the proof.}

\bibliographystyle{IEEEtran}
\phantomsection\addcontentsline{toc}{section}{\refname}\bibliography{multibandsensing}

\end{document}